\documentclass{aa}
\usepackage{graphicx}
\usepackage{txfonts}
\usepackage{float}
\usepackage{lipsum}
\usepackage{subcaption}
\usepackage{placeins}  
\usepackage{subcaption}
\usepackage{booktabs}
                                
\usepackage[colorlinks=true,linkcolor=blue, citecolor=blue, urlcolor=blue]{hyperref}

\renewcommand{\linenumbers}{}

\begin{document}

   \title{\texttt{GalaxyGenius}: Mock galaxy image generator for various telescopes from hydrodynamical simulations}

   \subtitle{}

    \author{Xingchen Zhou\inst{1,2}
    \and Hang Yang\inst{1}
    \and Nan Li\inst{1,2}\fnmsep\thanks{nan.li@nao.cas.cn}
    \and Qi Xiong\inst{1}
    \and Furen Deng\inst{1}
    \and Xian-Min Meng\inst{1}
    \and Renhao Ye\inst{3}
    \and Shiyin Shen\inst{3}
    \and Peng Wei\inst{1,2}
    \and Qifan Cui\inst{4}
    \and Zizhao He\inst{5}
    \and Ayodeji Ibitoye\inst{6,7,8}
    \and Chengliang Wei\inst{5}
    \and Yuedong Fang\inst{9}
    }

    \institute{National Astronomical Observatory, Chinese Acedemy of Science, 100101, Beijing, China 
    \and Science Center for China Space Station Telescope, National Astronomical Observatories, Chinese Academy of Sciences, 100101, Beijing, China
    \and Shanghai Astronomical Observatory, Chinese Academy of Sciences, 200030, Shanghai, China
    \and Shanghai Key Lab for Astrophysics, Shanghai Normal University, 200033, Shanghai, China
    \and Purple Mountain Observatory, Chinese Academy of Sciences, 210023, Jiangsu, China
    \and Department of Physics, Guangdong Technion – Israel Institute of Technology, 515063, Guangdong, China 
    \and Centre for Space Research, North-West University, Potchefstroom 2520, South Africa
    \and Department of Physics and Electronics, Adekunle Ajasin University, P. M. B. 001, Akungba-Akoko, Ondo State, Nigeria
    \and Universitäts-Sternwarte München, Fakultät für Physik, LudwigMaximilians-Universität München, Scheinerstrasse 1, 81679, München, Germany}

   \date{Received September 30, 20XX}

 
  \abstract
   {}
   {We introduce \texttt{GalaxyGenius}, a Python package designed to produce synthetic galaxy images tailored to different telescopes based on hydrodynamical simulations. Its implementation will support and advance research on galaxies in the era of large-scale sky surveys,
} 
   {The package comprises three main modules: data preprocessing, ideal data cube generation, and mock observation. Specifically, the preprocessing module extracts necessary properties of star and gas particles for a selected subhalo from hydrodynamical simulations and creates the execution file for the following radiative transfer procedure. Subsequently, building on the above information, {the ideal data cube generation module executes a widely used radiative transfer project, specifically the SKIRT}, to perform the SED assignment for each particle and the radiative transfer procedure to produce an IFU-like ideal data cube. Lastly, the mock observation module takes the ideal data cube and applies the throughputs of aiming telescopes, while also incorporating the relevant instrumental effects, point spread functions (PSFs), and background noise to generate the required mock observational images of galaxies.}
   {To showcase the outcomes of \texttt{GalaxyGenius}, we created a series of mock images of galaxies based on the IllustrisTNG and EAGLE simulations for both space and ground-based surveys, spanning   ultraviolet (UV) to infrared (IR) wavelength coverage, including CSST, Euclid, HST, JWST, Roman, and HSC.}
  {\texttt{GalaxyGenius} offers a flexible framework  to generate mock galaxy images with customizable recipes. These generated images can serve as valuable references for verifying and validating new approaches in astronomical research. They can also serve as training sets for relevant studies using deep learning in cases where real observational data are insufficient.}


    \titlerunning{\texttt{GalaxyGenius}}
    \authorrunning{X. Zhou et al.}
   \maketitle

\section{Introduction}\label{introduction}
Galaxies are the fundamental building blocks of the Universe, serving as cosmic laboratories where stars, planetary systems, and various astrophysical phenomena are born and evolve. In astronomy, the study of galaxies provides insights into the processes of star formation, stellar dynamics, and the interstellar medium (ISM), enhancing our understanding of the life cycle of celestial objects~\citep{Madau2014,Conselice2014,Behroozi2013,Kravtsov2012}. In cosmology, galaxies are crucial for mapping the large-scale structure of the Universe, offering clues on the distribution of dark matter and the influence of dark energy on cosmic expansion~\citep{Wang2020universal,Wang2020brief, Springel2006,Weinberg2013}. Observations of galaxies at different redshift allow scientists to look back in time and study the Universe's evolution, shedding light on the formation and growth of cosmic structures over billions of years. 

In current astronomical and cosmological studies, simulations play a pivotal role, which can bridge the gap between theoretical models and observational data~\citep{Schaye2015,Sharbaf2024,Harnois2018,Villaescusa2020}. Galaxy simulations are essential for advancing our understanding of the Universe. They allow researchers to model complex processes such as galaxy formation, evolution, and interactions, which are difficult to study through direct observation alone. By incorporating various physical laws and parameters, these simulations help scientists test hypotheses, explore parameter spaces, and predict observable signatures. Additionally, the performance of instruments, data analysis pipeline and algorithms used in large-scale photometric and spectroscopic surveys can be comprehensively investigated. 

Currently, the most sophisticated and realistic galaxies are simulated or generated from hydrodynamical simulations. These categories of simulations are distinct from N-body simulations, which only consider the evolution of dark matter. Instead, hydrodynamical simulations are computational tools that model the behavior of fluids, such as gases and liquids, under various physical conditions. In astrophysics, hydrodynamical simulations are crucial for understanding complex phenomena like galaxy formation, star formation, and the dynamics of interstellar and intergalactic media~\citep{Nelson2019illustris,Genel2014}. By solving the equations of fluid dynamics, often coupled with additional physics like gravity, magnetic fields, gas cooling and feedback from supernovae and active galactic nuclei (AGNs) and radiation, researchers can recreate and study the evolution of cosmic structures over time and reproduce many observed properties of galaxies~\citep{Crain2023,Valentini2025}. They help elucidate the role of feedback mechanisms in regulating star formation and shaping the ISM. Additionally, by comparing simulation results with observational data, researchers can test and refine theoretical models, leading to a more comprehensive understanding of the universe's history and structure. Utill now, hydrodynamical simulations, such as Illustris~\citep{Vogelsberger2014, Genel2014, Sijacki2015}, IllustrisTNG~\citep{Marinacci2018, Pillepich2018, Pillepich2019, Nelson2018, Nelson2019first, Naiman2018, Springel2018}, EAGLE~\citep{eagle2017simulations, McAlpine2016}, FLAMINGO~\citep{Schaye2023}, and other simulations, are widely utilized to address the challenges in galaxy formation and evolution and promote our understanding of our Universe. 

From hydrodynamical simulations, galaxies can be generated by assigning particles in a subhalo spectral energy distributions (SEDs) based on particle properties. {To account for the dust effects, radiative transfer processes or simplified prescriptions shown in~\citet{Martin2022, Fortuni2023, LaChance2025} can be conducted to model the absorption, scattering, and emissions.} This process typically generates three-dimensional (3D) ideal data cubes. Subsequently, the data cubes are convolved with throughputs of filters for specific survey to obtain the bandpass images. Finally, point spread function (PSF) and instrumental effects are applied to generate mock galaxy images. Several endeavors exist for generating galaxies from the EAGLE~\citep{Crain2015, Schaye2015} and Illustris~\citep{Vogelsberger2014, Genel2014, Sijacki2015} simulations, {as well as its upgraded version}, IllustrisTNG~\citep{Marinacci2018, Pillepich2018, Pillepich2019, Nelson2018, Nelson2019first, Naiman2018, Springel2018}, {employing radiative transfer codes}, such as DIRTY~\citep{Gordon2001}, SKIRT~\citep{Baes2011, Camps2015}, HYPERION~\citep{Robitaille2011}, RADMC-3D~\citep{Dullemond2012}, POLARIS~\citep{Reissl2016}, and SOC~\citep{Juvela2019}. 

{These generated galaxies serve as a versatile platform for extensive applications in galaxy surveys. One such application is machine learning-based studies, particularly those utilizing deep learning techniques}, such as the estimation of photometric redshifts~\citep{Naidoo2023,Moskowitz2024,Zhou2022Extracting, Zhou2022Photometric}, classification of galaxy morphologies~\citep{Pfeffer2023, Hambleton2011, Dickinson2018, Gong2025mock}, measurement of galaxy properties (shear, stellar mass)~\citep{Csize2024,Tewes2019,Zhang2024,Bonjean2019,Chu2024, Tang2018investigation, Tang2020satellite, Tang2021importance, Tang2023mock}, {identification of galaxy mergers}~\citep{Ferreira2020,Pearson2019,Nevin2019,Snyder2019,Bottrell2019, Guzman2023, Rose2023, Margalef2024,Wilkinson2024}, and investigation of blending effects and deblending algorithms~\citep{Arcelin2021, Burke2019, Hemmati2022}. These galaxies can effectively address the insufficiency of training data in these studies and can aid in the development of a robust deep learning model. Moreover, these galaxies provide a controlled and well-defined environment where ground-truth information is precisely known. This allows us to test, validate, and calibrate the performance of deep learning models under various observational conditions, such as noise, resolution, and PSF effects. Furthermore, by generating mock observations that mimic real survey data, we can assess how well models generalize to real-world datasets.

{Several projects are dedicated to generating mock observational galaxy images from hydrodynamical simulations, such as FORECAST~\citep{Fortuni2023} and RealSim~\citep{Bottrell2017galaxies,Bottrell2019}. These projects employed simplified prescriptions, including semi-analytic models~\citep{Guiderdoni1987, Devriendt2000,Nelson2019illustris,Vogelsberger2020}, to account for absorption, scattering, and re-emission of dust components. While in our work, we present a Python package, \texttt{GalaxyGenius}, that is capable of generating galaxies in ultraviolet (UV) to infrared (IR) bands, incorporating comprehensive radiative transfer procedures implemented by SKIRT. Although the radiative transfer process is computationally intensive, this approach generates galaxies with enhanced physical details and realism, as it self-consistently models the interplay between radiation and interstellar dust.}
This package integrates three modules: data preprocessing, ideal data cube generation, and mock observation. The preprocessing encompasses extraction of particles and corresponding properties used for SED assignment from hydrodynamical simulations {and creation of an execution file for the radiative transfer code.} Subsequently, data cube generation module produces an IFU-like data by executing SKIRT based on information obtained from preprocessing. Finally, mock observation module generate realistic observational images by incorporating instrumental effects specific for photometric instruments. We demonstrate the capability of \texttt{GalaxyGenius} by creating mock images from IllustrisTNG and EAGLE simulations for both space and ground-based surveys spanning  UV to IR wavelength coverage, including the Chinese Space Station Telescope (CSST, ~\citet{Zhan2018,Gong2019}), Hubble Space Telescope (HST, ~\citet{Freedman2001,Momcheva2016}), James Webb Space Telescope (JWST, ~\citet{Sabelhaus2004}), Euclid Space Telescope (Euclid, ~\citet{Mellier2024}), Nancy Grace Roman Space Telescope (Roman,~\citet{Spergel2015}), and Hyper Suprime-Cam (HSC, \citet{Aihara2018}). Additionally, we also investigate the effects of dust elements on galaxy generation, particularly for dust recipes and dust models. Furthermore, we discuss some limitations on generating galaxies from hydrodynamical simulations using SKIRT project. 

This paper is organized as follows. Section~\ref{sec: GalaxyGenius} elaborates the implementation details of \texttt{GalaxyGenius} on generating galaxies from hydrodynamical simulations using SKIRT. We demonstrate some sample outputs for various photometric surveys in Section~\ref{sec: sample outputs}. Subsequently, in Section~\ref{sec: discussion}, we discuss the effects of dust recipes and dust models and explain some limitations of generating galaxies from hydrodynamical simulations. Finally, this work is summarized in Section~\ref{sec: conclusion}. {Appendix~\ref{sec: instrumental parameters} describe the instrumental parameters for each considered survey and their origins.} Appendix~\ref{app: psfs} displays PSFs utilized for photometric bands for each considered survey. {Appendix~\ref{app: derivation} provides the derivation details for noise levels achieved from limiting magnitudes.}

\section{\texttt{GalaxyGenius}}\label{sec: GalaxyGenius}

In this section, we initially describe the overall framework of \texttt{GalaxyGenius}. Subsequently, we provide a detailed explanation of the entire routine, including data preprocessing, ideal data cube generation, and mock observations for generating galaxy images utilizing IllustrisTNG simulations as a case study. 

\subsection{Overall framework}\label{sec: overall framework}
\begin{figure*}
    \centering
    \includegraphics[width=1.\linewidth]{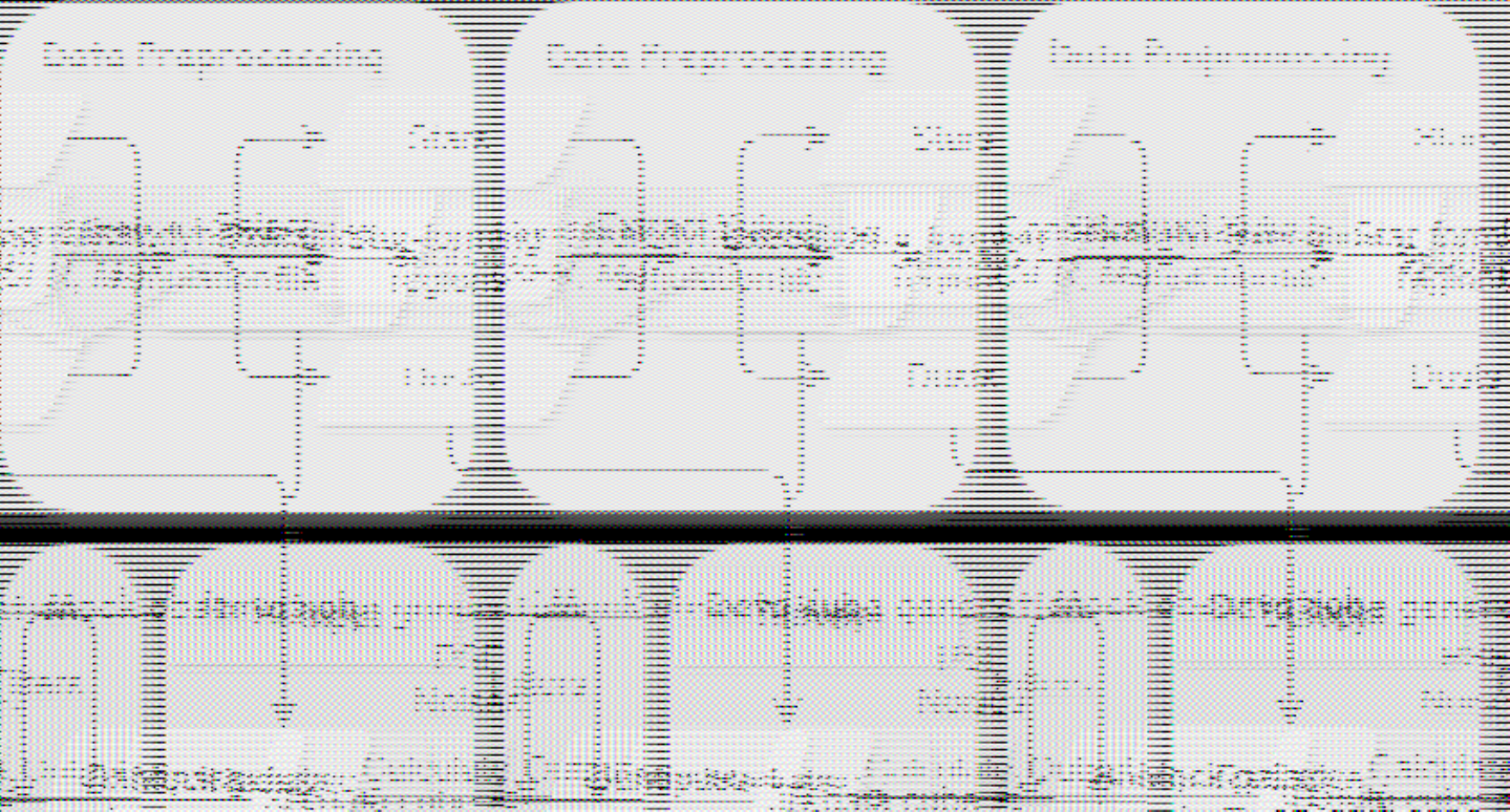}
    \caption{Overall framework for \texttt{GalaxyGenius}. This framework encompasses three primary modules: data preprocessing, data cube generation, and mock observation. The data preprocessing selects particular subhalo under certain criteria, from where particle and execution files are created. Data cube generation run radiative transfer from the output previously to produce 3D data cube and SEDs. Finally, mock observation constructs mock galaxy images based on specific survey.}
    \label{fig: overall framework}
\end{figure*}

The overall framework of \texttt{GalaxyGenius} is illustrated in Figure~\ref{fig: overall framework}, which comprises three primary modules for generating mock observational galaxy images tailored to specific surveys, described below.

    (i) Data preprocessing: this module extracts coordinates and relevant properties for different categories of particles, including stars, star-forming regions and dust elements, {for a selected subhalo of hydrodynamical simulation.} Additionally, it also prepares the execution file for radiative transfer codes based on the particle files and configurations. For further details, we refer to Section~\ref{sec: preprocess}.

    (ii) Ideal data cube generation: this module executes the radiative transfer procedures based on the particle files and execution file created in data preprocessing. Finally, an IFU-like ideal data cubes and SEDs for different viewing angles can be obtained. We refer to Section~\ref{sec: datacube gen} for more details.

    (iii) Mock observation: this module aims to construct mock galaxy images anticipated for specific survey. It initially creates ideal bandpass images from the data cube outputted above, utilizing the throughput of each filter. Subsequently, it applies instrumental effects, including point spread functions (PSF) and background and instrumental noises under specified observational conditions. The details are described in Section~\ref{sec: postprocess}.

\subsection{Data preprocessing}\label{sec: preprocess}
\renewcommand{\arraystretch}{1.2}
\newcommand{\wraptext}[2]{\parbox[t]{#1}{\raggedright #2}}
\captionsetup{font=normalsize}
\begin{table}[h!]
\caption{Necessary properties for stars, star-forming regions, and dust elements and their origins.}
\label{tab: parameters}
\centering
\begin{tabular}{p{1.5cm}p{2cm}p{4cm}}
\hline
Parameters     & Origin      & Description                                                \\ \hline
\multicolumn{3}{c}{\textbf{Stars}}                                                        \\
$x, y, z$      & Simulation  & Coordinates                                                \\
$h$            & Simulation  & Smoothing length                                           \\
$M_{\rm init}$ & Simulation  & Initial mass                                               \\
$Z$            & Simulation  & Metallicity                                                \\
$t$            & Simulation  & Age                                                        \\ \hline
\multicolumn{3}{c}{\textbf{Star-forming regions}}                                        \\
$x, y, z$      & Simulation  & Coordinates
\\
$h$            & Simulation  & Smoothing length                                           \\
SFR            & Calculated  & \wraptext{5cm}{SFR of the HII region}                     \\
$Z$            & Simulation  & Metallicity of the HII region                              \\
$C$            & Free param. & Compactness of the HII region                              \\
$P$            & Free param. & Pressure of the ambient ISM                                \\
$f_{\rm PDR}$  & Free param. & \wraptext{5cm}{Dust covering factor of the photodissociation region (PDR)} \\ \hline
\multicolumn{3}{c}{\textbf{Dust}}                                                         \\
$x, y, z$      & Simulation  & Coordinates
\\
$h\ ^{a}$            & Simulation  & Smoothing length                                           \\
$M$            & Simulation  & Mass of gas particles                                      \\
$Z$            & Simulation  & Metallicity                                                \\ 
$T$            & Calculated  & Temperature                                                \\ \hline
\end{tabular}
\\
\vspace{1mm}
\begin{flushleft}
{$^a$ Only include for simulations conducted by SPH method for gases. }
\end{flushleft}
\end{table}

The data preprocessing module of \texttt{GalaxyGenius} is utilized to extract relevant particles from a particular subhalo obtained from hydrodynamical simulations. Users have the option to prepare the particles in advance. In such a scenario, several crucial parameters must be provided as input to data preprocessing module through an interface to construct the execution file for radiative transfer in subsequent step. These parameters encompasses subhaloID, redshift, cosmology, stellar mass, and the box length for particle selection. Given that IllustrisTNG is a suite of state-of-the-art magneto-hydrodynamical simulations and is widely employed to generate galaxies, we provide a specialized interface for data preprocessing tailored for this set of simulations. The following paragraphs provide a detailed explanation of the required particle files, utilizing IllustrisTNG as an illustrative example. 

Here, we take snapshot-94 ($z\approx0.06$) of TNG100 from IllustrisTNG as a case study. Firstly, subhalos are read from the subhalo catalog derived using the Subfind algorithm~\citep{Springel2000GADGETAC}. The subhalos in the desired stellar mass range should have a constraint as \textsc{SubhaloFlag == 1}, which is to ensure that these subhalos can be considered as galaxy or satellite of cosmological origin~\citep{Rios2023, Nelson2019illustris}. 

\renewcommand{\arraystretch}{1.2}
\begin{table}
\caption{Stellar masses $M_{\star}$, SFR, number of particles represented by BC03, MAPPINGS-III SED families, and number of dust elements derived from two recipes for three example subhalos (with IDs of 0, 31, and 253881.)}
\label{tab: metadata}
\begin{tabular}{|l|c|c|c|}
\hline
SubhaloID                  & 0            & 31           & 253881       \\ \hline\hline
$M_{\star}$ ($M_{\odot}$)  & $10^{12.55}$ & $10^{10.59}$ & $10^{11.63}$ \\ \hline
SFR ($M_{\odot}/yr$)       & 1.89         & 14.03        & 14.40        \\ \hline
$N_{\rm BC03}$              & 2,814,159    & 38,823       & 432,621      \\ \hline
$N_{\rm MAPPINGS-III}$       & 8            & 308          & 276          \\ \hline
$N_{\rm DUST, Camps+2016}$  & 740          & 10,080       & 19,651       \\ \hline
$N_{\rm DUST, Torrey+2012}$ & 1,736        & 13,404       & 42,047       \\ \hline
\end{tabular}
\end{table}

Secondly, the star particles in one subhalo are retrieved within a cubic box as radiation sources. Particles with star formation time higher than 0 are selected to exclude the wind particles. The boxsize is defined as 20 times half stellar mass radius $R_{\rm 1/2, \star}$, with a maximum value of 300 kpc to limit the spatial pixels of output data cubes~\citep{Bottrell2024}. The coordinates of each particle are read directly from the snapshot, while the smoothing length is assigned based on redshift of the snapshot, considering that the smoothing length is 0.74 kpc in physical scale at $z=0$~\citep{Nelson2019illustris}. 

\begin{figure*}
    \centering
    \includegraphics[width=1.0\linewidth]{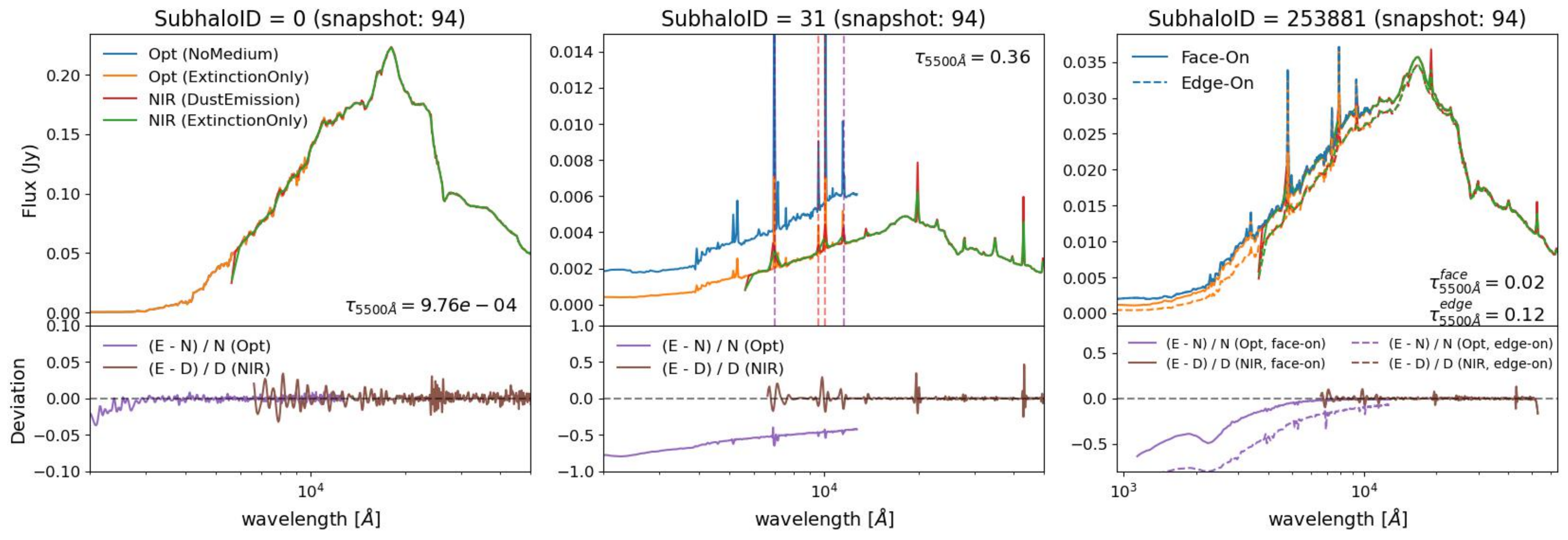}
    \caption{{Ideal SEDs for subhalo 0, 31, and 253881 output by SKIRT. For comparison, NoMedium and ExtinctionOnly mode are employed to generate the optical SEDs, while the NIR ones are generated using DustEmission and ExtinctionOnly mode. The relative deviations in optical and NIR are displayed in the lower panel, where N, E, and D represent NoMedium, ExtinctionOnly, and DustEmission. The optical depth at 5500{\AA} $\tau_{5500\text{\AA}}$ are calculated and shown in each panel. The viewing angles of subhalo 0 and 31 are both at $0^\circ$ and $0^\circ$ for inclination and azimuth, while subhalo 253881 is in two views, face-on and edge-on, indicated in solid and dashed lines respectively. For subhalo 31, we also highlight several emission lines, H$\alpha$ 6563, [SIII]$\lambda\lambda\ 9069,9532$, and Pa$\gamma$ 10941 from left to right. These lines are generated by star-forming regions represented by MAPPINGS-III SED family.}}
    \label{fig: seds}
\end{figure*}

We follow the procedure established by~\citet{Rodriguez2019}. For each particle older than 10 Myr, stellar population SED from the Bruzual \& Charlot (BC03, ~\citet{Bruzual2003}) or Flexible Stellar Population Synthesis (FSPS, \citet{Conroy2010}) can be employed. Here, BC03 SSP family with Chabrier initial mass function~\citep{Chabrier2003} is assigned. We call them as stars or BC03 particles interchangeably in following discussions. The initial mass, $M_{\rm init}$, metallicity, $Z$, and age, $t$, for the SSP are directly inherited from TNG data. 

While for particles younger than 10 Myr, it is assumed that they are still partially embedded within their respective clouds. Consequently, these particles are assigned SEDs based on a library of HII region templates derived from the MAPPINGS-III library~\citep{Groves2008}. We call these particles as star-forming regions in the following context. Each particle's SED is characterized by five parameters. Among these parameters, metallicity, $Z$, can be directly retrieved from particle data. The SFR is assumed to remain constant throughout HII region's lifetime of 10 Myr. On the other hand, the remaining three parameters are free ones. The compactness, $C$, determines the dust temperature distribution and, consequently, it influences the shape of far-IR (FIR) continuum dust emission. We sample the compactness using a lognormal Gaussian distribution with a mean of $\log C = 5$ and a standard deviation of 0.4 for each particle~\citep{Kapoor2021,Trcka2022}. The ISM pressure $P$ only affects the strength of several emission lines but does not alter the shape of the SED~\citep{Groves2008}. Therefore, we treat it as a constant value of $P = 1.38\times10^{-12}\ \rm Pa$. The final one, covering factor of t\textit{}he photodissociation region (PDR) $f_{\rm PDR}$ describes the extent that HII regions are coated~\citep{Kaufman1999, Pound2023}. Since the coating will eventually clear in PDR, this parameter is set as $f_{\rm PDR} = \exp(-t/\tau_{\rm clear})$, where $t$ represents the age of the stellar particle, and $\tau_{\rm clear}$ is the molecular cloud clearing or dissipation time scale. $\tau_{\rm clear} = 0$ means that this star-forming region is thoroughly transparent, while $\tau_{\rm clear} = \inf$ indicates that it is completely covered. We adopt the value of 3 Myr, as recommended in~\citet{Trcka2022}.

The next step involves modeling the distribution of dust elements within each galaxy. The density of dust at any given volume is determined by the assumption that a predetermined percentage of the metals present in the ISM gas are encapsulated within dust grains, as follows:
\begin{equation}
\rho_{\rm dust} =
\begin{cases} 
f_{\rm dust}Z_{\rm gas}\rho_{\rm gas}, & \text{if ISM}, \\
0, & \text{otherwise},
\end{cases}
\end{equation}
where $Z_{\rm gas}$ and $\rho_{\rm gas}$ represents the metallicity and density of the gas, respectively. Also, $f_{\rm dust}$ denotes the dust-to-metal ratio. This ratio is correlated with stellar mass, according to work by~\citet{Lu2023}, however, we assume it to be a constant of 0.3 for simplicity. Furthermore, we distinguish between the hot circumgalactic medium (CGM) and the cooler ISM gas using two distinct dust recipes:

(i) Only gas cells with non-zero SFR or with a temperature, $T$, below $8000K$ are considered as ISM~\citep{Camps2016}:
\begin{equation}\label{eq: camps}
    {\rm SFR} > 0\ {\rm or}\ T < 8000K
,\end{equation}
where $T$ is calculated from internal energy, $u$, and electron abundance, $x_e$, as recommended in TNG website~\footnote{\url{https://www.tng-project.org/data/docs/faq/\#gen6}} as follows:
\begin{equation}\label{eq: temperature}
    T = (\gamma - 1) \times \left(\frac{u}{k_B}\right) \times \left(\frac{4}{1 + 3X_H + 4X_H x_e} \times m_p\right)
,\end{equation}
where $\gamma=5/3$, $k_B$, $X_H$, and $m_p$ are the adiabatic index, the Bolzmann constant, the hydrogen mass fraction, and the mass of proton respectively. The condition ${\rm SFR} > 0$ captures gas particles that are eligible for star formation, but were not actually converted into a star-forming region. Meanwhile, the temperature cutoff value of $8000K$ for non-star-forming gas particles ensures that only gas capable of forming dust is considered, as dust cannot form or is rapidly destroyed in hot gas (\citep{Camps2016, Guhathakurta1989}).
The cutoff temperature is arbitrary and empirical. However, this condition does not cause a substantial difference in the dust geometry {(see Section~\ref{sec: dusts})}, since the diffuse dust content of each galaxy is essentially determined by the star-forming gas alone~\citep{Rodriguez2019, Kapoor2021}.

(ii) The rotationally supported interstellar gas, settled in the disc, are distinguished from the hot CGM gas following~\citet{Torrey2012}:
\begin{equation}
    \log\left(\frac{T}{K}\right) < 6 + 0.25\log\left(\frac{\rho_{\rm gas}}{10^{10}\ h^2\ M_{\odot}\ {\rm kpc^{-3}}}\right)
,\end{equation}
where $T$ is temperature calculated using Equation~\ref{eq: temperature} and $\rho_{\rm gas}$ is gas density directly obtained from snapshot data. This condition has been shown to effectively remove cells in the hot halo~\citep{Torrey2012,Mckinnon2016}. To enhance the adaptability of \texttt{GalaxyGenius}, we implement both dust recipes, allowing users to freely choose between them. {Furthermore, in TNG, hydrodynamics for the gas component is solved using a moving Voronoi mesh via the AREPO code~\citep{Weinberger2020}, eliminating the need for smoothing length.} The necessary properties for stars, star-forming regions, and dust elements along with their origins are summarized in Table~\ref{tab: parameters}. 

We employed SKIRT to perform the radiative transfer procedures. SKIRT is a versatile and widely used Monte Carlo (MC) radiative transfer code~\footnote{\url{https://skirt.ugent.be/}} designed to model the interaction of light with matter in astrophysical systems~\citep{Baes2011,Camps2015}. Developed to study processes such as dust absorption, scattering, and thermal re-emission, SKIRT is capable of generating realistic synthetic observations of galaxies, star-forming regions, and other celestial objects. A key feature of SKIRT is its ability to incorporate complex geometries, spatial distributions of stars and dust, and wavelength-dependent properties of materials. Through MC techniques, SKIRT simulates the paths of photon packets as they propagate through a medium, accounting for scattering, absorption, and re-emission processes. SKIRT is particularly well-suited for creating mock observations from hydrodynamical simulations, such as IllustrisTNG and EAGLE~\citep{Nelson2018, Rodriguez2019,Shen2020,Camps2016,Guzman2023,Rios2023,Bottrell2024,Baes2024}. In such scenario, SKIRT transforms theoretical outputs, such as stellar population distributions and gas content, into observable quantities, including images and SEDs across various wavelength coverage, from UV to FIR. The execution file for SKIRT, with a $\rm .ski$ extension, can be generated through a user-friendly command-line Q\&A session. 

{Several critical parameters influence the output of SKIRT.} {These parameters include simulation mode, dust model, wavelength range, number of photon packets, and some instrument-specific configurations such as field of view, number of spatial pixels, observing directions, and distance.} {The commonly utilized simulation modes include NoMedium, ExtinctionOnly, and DustEmission. These modes represent ideal scenarios with no dust, concentrate solely on extinction and scattering effects, and consider extinction and scattering along with dust emission, respectively. Given that dust emission primarily affects the IR wavelength range~\citep{Draine2003,Meny2007,daCunha2008,Massa2014,Decleir2022}, we employ ExtinctionOnly and DustEmission modes for optical and IR ranges, respectively. The NoMedium mode is preserved for testing purposes due to its rapid execution by considering the emissions solely from primary sources.} {For modeling dust grains, SKIRT offers several options, including the dust model introduced in~\citet{Zubko2004} (ZubkoDustMix),~\citet{Draine2007} (DraineLiDustMix), and~\citet{Jones2017} (ThemisDustMix).} We adopted a fixed dust grain model, ZubkoDustMix, in every location of the galaxy. This model comprises three dust components: silicate, graphite, and polycyclic aromatic hydrocarbons (PAHs), with size distributions defined in 15 bins. {We employed hierarchical octree with refinement level from 6 to 10 to re-grid the dust density to increase the speed of radiative transfer for dust, instead of using the native Voronoi grid in TNG simulation~\citep{Baes2024}.} Further details on other dust models and their effects on generating galaxies are provided in Section~\ref{sec: dusts}. {The wavelength range should be set to encompass the wavelength of all the considered filters, taking into account the effect of redshift.} The number of photon packets in MC simulation plays a crucial role in determining the signal-to-noise ratio (S/N) of the output and the computational cost. To strike a balance between S/N and efficiency, we used a fixed value of $10^8$ photon packets for all simulations. This choice ensures high-quality results while maintaining reasonable computational demands. {Regarding instrument-specific configurations, the field of view can be freely adjusted, depending upon the aperture size of considered instrument. Consequently, the outputs including data cubes and SEDs will be limited to this specific size. The number of spatial pixels should be set to correspond to the pixel scale of the specific filter. Otherwise, rescaling is necessary to generate mock images. It is advisable to set observing directions with caution, as the memory usage significantly increases with the number of observing directions. Observing distance can be calculated based on redshift if a flat universe is assumed to account for the cosmological surface brightness dimming. In contrast, for a local universe, the distance should be explicitly specified.}

The snapshot and subhalo data of IllustrisTNG, from which the particles were retrieved, are of an exceptionally large size. Fortunately, a web-based interface (API)~\footnote{\url{https://www.tng-project.org/data/docs/api/}} can be utilized to search, extract, visualize, and analyze data without the necessity of acquiring the entire simulation data. The utilization of this web-API was also implemented in \texttt{GalaxyGenius} to facilitate galaxy generation. 

\subsection{Data cube generation}\label{sec: datacube gen}
The radiative transfer procedure is executed using the particle files and the SKIRT execution file generated during the preprocessing stage mentioned above. This process assigns each particle a SED based on its properties and simulates the propagation of light through the ISM of the galaxy, accounting for absorption, scattering, and emission. {The memory consumption during this step is primarily influenced by two factors: the spatial and wavelength resolution and the number of viewing angles. Higher resolution and more viewing angles increase the computational and memory requirements significantly.}

SKIRT offers the capability to record outputs from individual radiative transfer components, providing detailed insights into the contributions of different physical processes. These components include transparent, primary direct, primary scattered, secondary transparent, secondary direct, secondary scattered, and total components. While recording these components can provide valuable diagnostic information, they consume a substantial amount of memory and storage. In practice, to optimize computational efficiency and reduce memory overhead, we typically record only the total components, which represent the aggregate effect of all radiative processes.

Upon completion of the radiative transfer procedure, the output includes an IFU-like ideal data cube with a shape of $(N_{\rm wavelength}, N_{\rm pixel}, N_{\rm pixel})$ {corresponding to the resolution setting described in Section~\ref{sec: preprocess}.} This data cube contains the flux values at each wavelength and spatial pixel, providing a comprehensive view of the galaxy's spectral and spatial properties. The flux units in the data cube are expressed in $\rm MJy/sr$. Additionally, SKIRT generates SEDs observed from the viewing angles specified in the execution file. These SEDs are presented in units of $\rm Jy\ versus\ micron$. The SEDs capture the integrated light from the galaxy as a function of wavelength, offering insights into the galaxy's stellar populations, dust properties, and overall energy distribution.

\subsection{Mock observation}\label{sec: postprocess}
After acquiring the data cubes, we can calculate the bandpass images, given the throughput of each filter for a particular instrument. All calculations are performed in units of electron counts, enabling straightforward incorporation of PSF effects and instrumental and background noises. The electron counts for each pixel can be determined by providing the exposure time, $t_{\rm exp}$, the number of exposures, $N_{\rm exp}$, the effective area of the telescope, $A_{\rm aper}$, and the pixel scale of the CCD camera, $l_p$, as follows {(see Section 9.2 in~\citet{Ryon2023})}:
\begin{equation}
    C = \frac{t_{\rm exp}N_{\rm exp}A_{\rm aper}l_p^2}{hc}\int_{\lambda_{\rm min}}^{\lambda_{\rm max}}\lambda f(\lambda)T(\lambda)d\lambda,
\end{equation}
where $h$ and $c$ are Planck constant and the speed of light respectively; then, $T(\lambda)$ is the throughput of filter and $f(\lambda)$ is flux measured per wavelength, converted from flux measured per frequency in units of $\rm MJy/sr$ employed by SKIRT using the following equation:
\begin{equation}\label{eq: nu_to_lambda}
    f_\lambda = f_\nu\frac{c}{\lambda^2}.
\end{equation}

\begin{figure*}
    \centering
    \includegraphics[height=9cm]{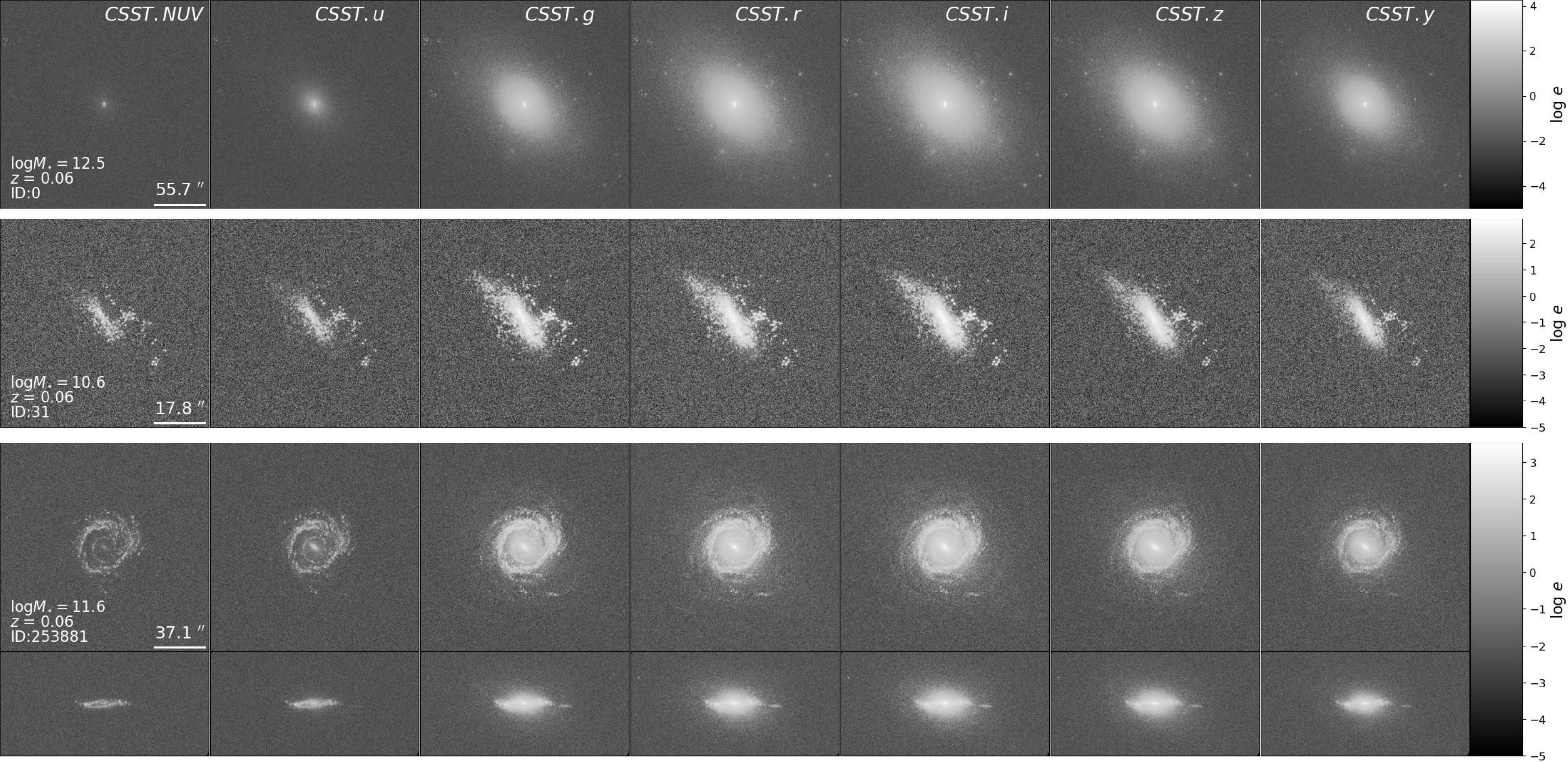}
    \caption{
    Mock galaxy images in $NUV,u,g,r,i,z$, and $y$ bands expected by CSST, {with the colorbar showing the units as electron counts.} The viewing angles are set to be $0^\circ$ and $0^\circ$ in inclination and azimuth for subhalos 0 and 31, while for subhalo 253881, the viewing angles are set to be at face-on and edge-on directions. {For better illustration, the spiral galaxy is zoomed into a FoV with 200 kpc,} and the edge-on images are rotated to make the galactic plane parallel to $x$-axis. Furthermore, the stellar mass, redshifts, and subhaloIDs are also displayed. The images are presented in logarithmic scale, and to mitigate the occurence of NaN values and enhance visual clarity, they are clipped at $10^{-5}$.}
    \label{fig: csst}
\end{figure*}

\begin{figure*}
    \centering
    \includegraphics[height=9cm]{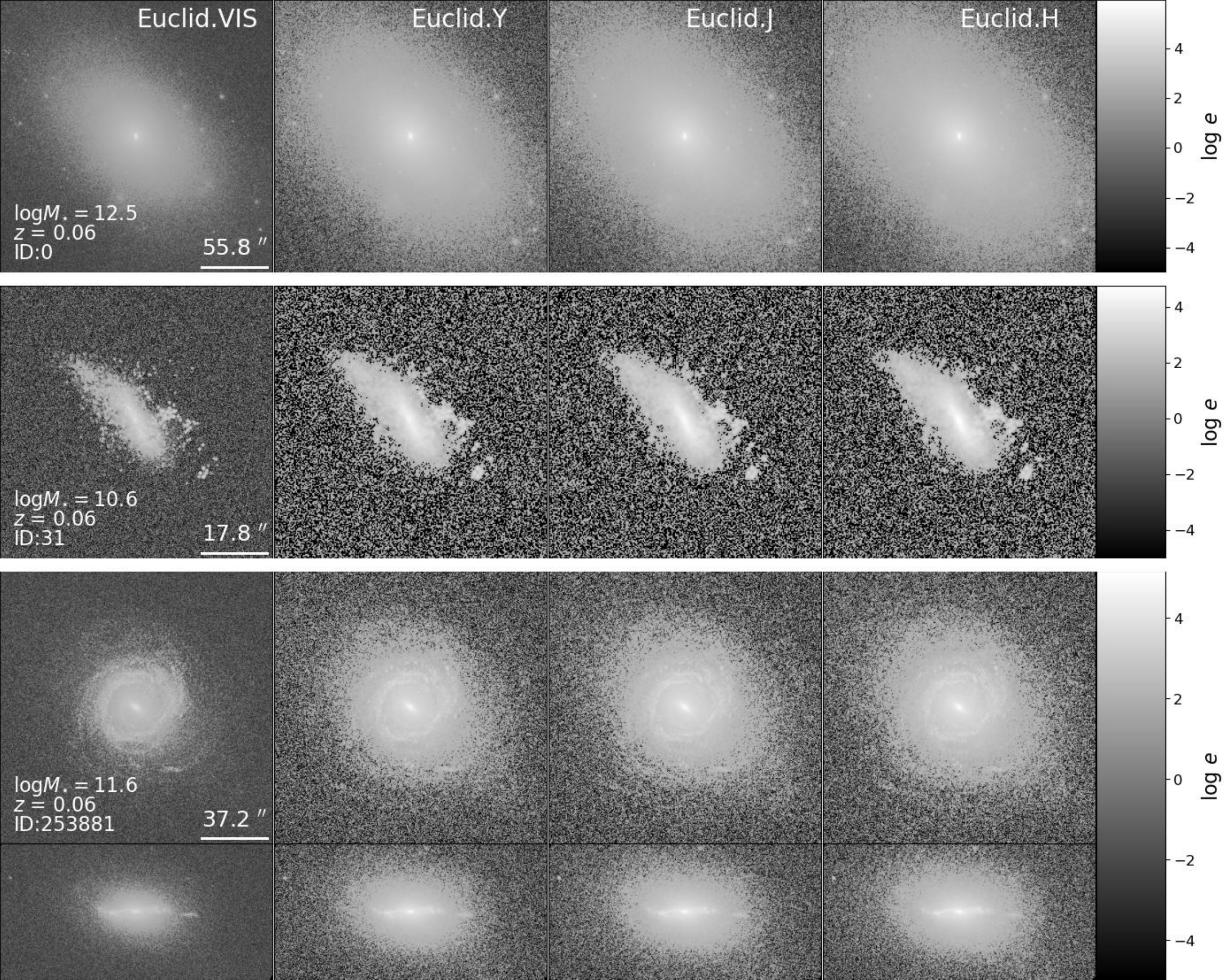}
    \caption{Mock galaxy images in VIS, Y, J, and H bands for Euclid.}
    \label{fig: euclid}
\end{figure*}

\begin{figure*}
    \centering
    \includegraphics[height=9cm]{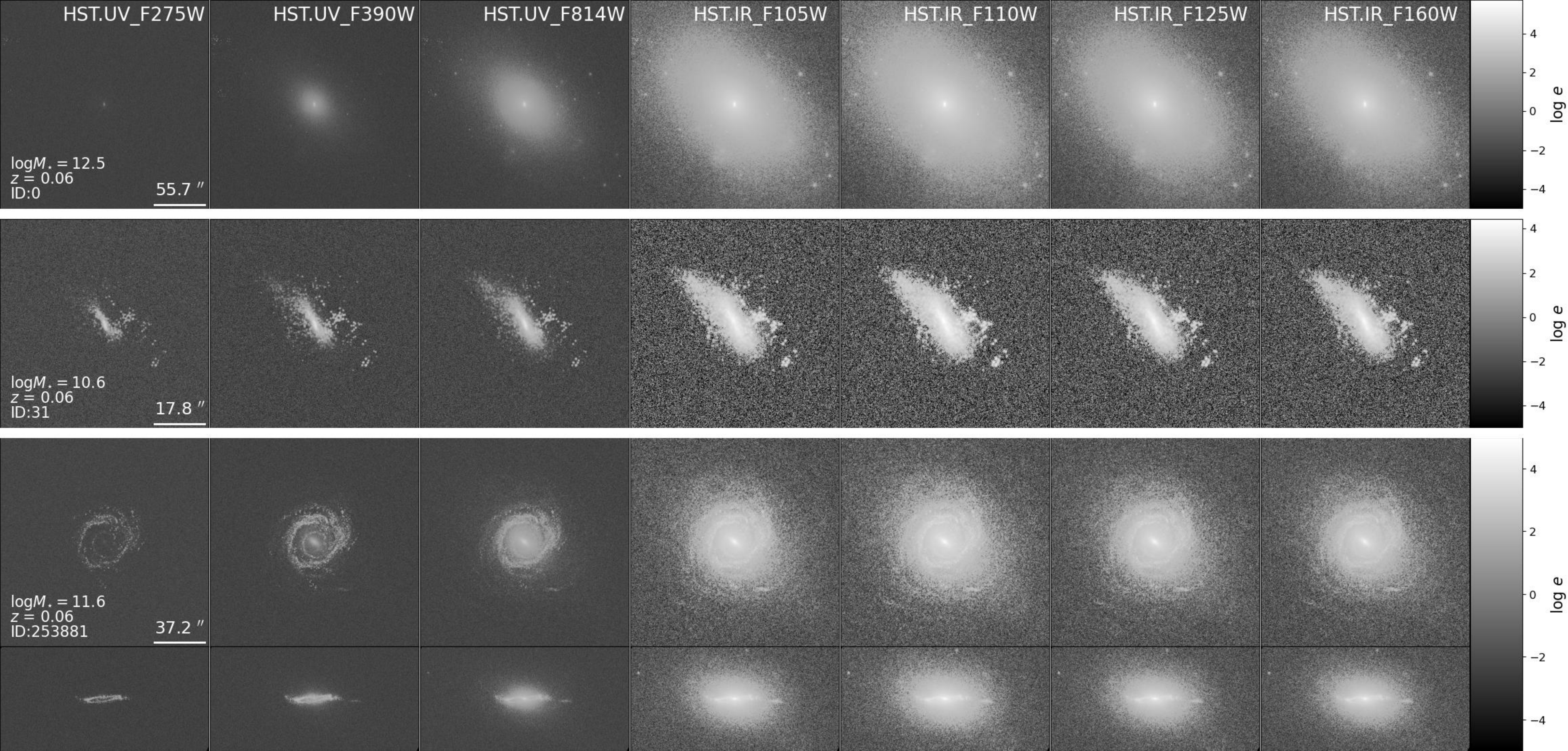}
    \caption{Mock galaxy images in F275W, F390W, and F814W bands from UVIS channel and F105W, F110W, F125W, and F160W bands from IR channel for HST WFC3.}
    \label{fig: hst}
\end{figure*}

\begin{figure*}
    \centering
    \includegraphics[height=9cm]{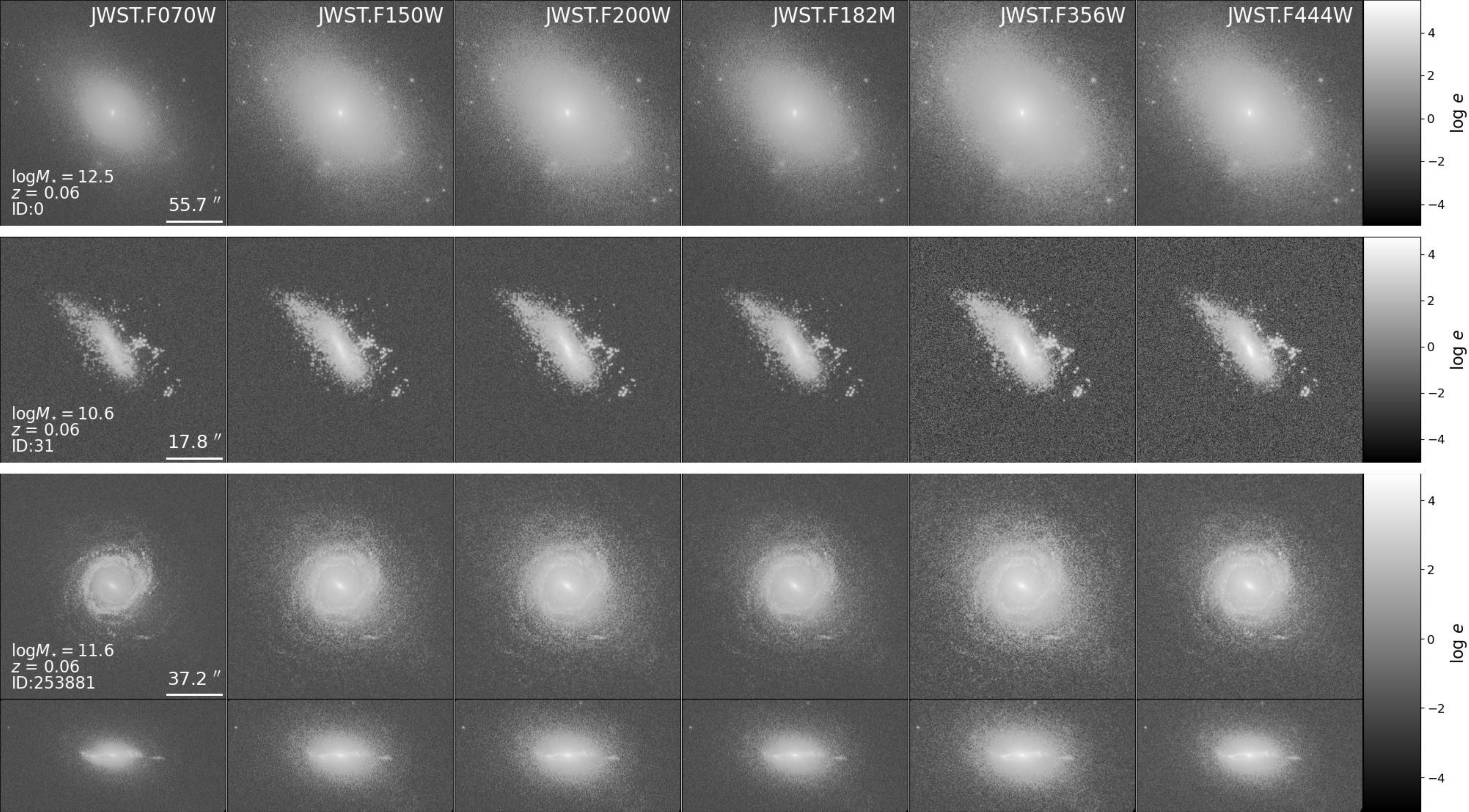}
    \caption{Mock galaxy images in F070W, F150W, F200W, and F182M bands from short-wavelength channel and F356W and F444W bands from the long-wavelength channel for JWST NIRCam.}
    \label{fig: jwst}
\end{figure*}

\begin{figure*}
    \centering
    \includegraphics[height=9cm]{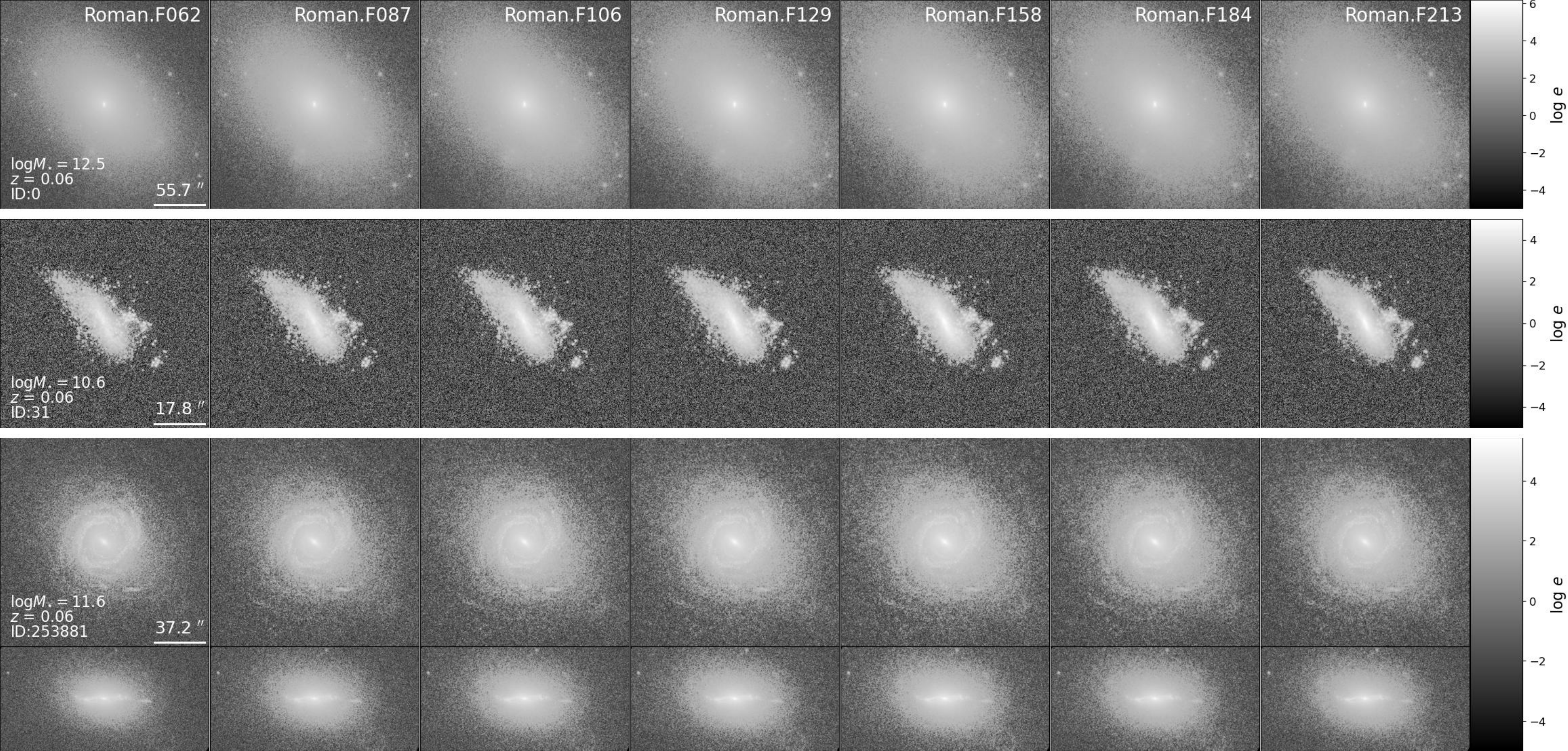}
    \caption{Mock galaxy images in F062, F087, F106, F129, F159, F184, and F213 bands for Roman WFI.}
    \label{fig: roman}
\end{figure*}

\begin{figure*}
    \centering
    \includegraphics[height=9cm]{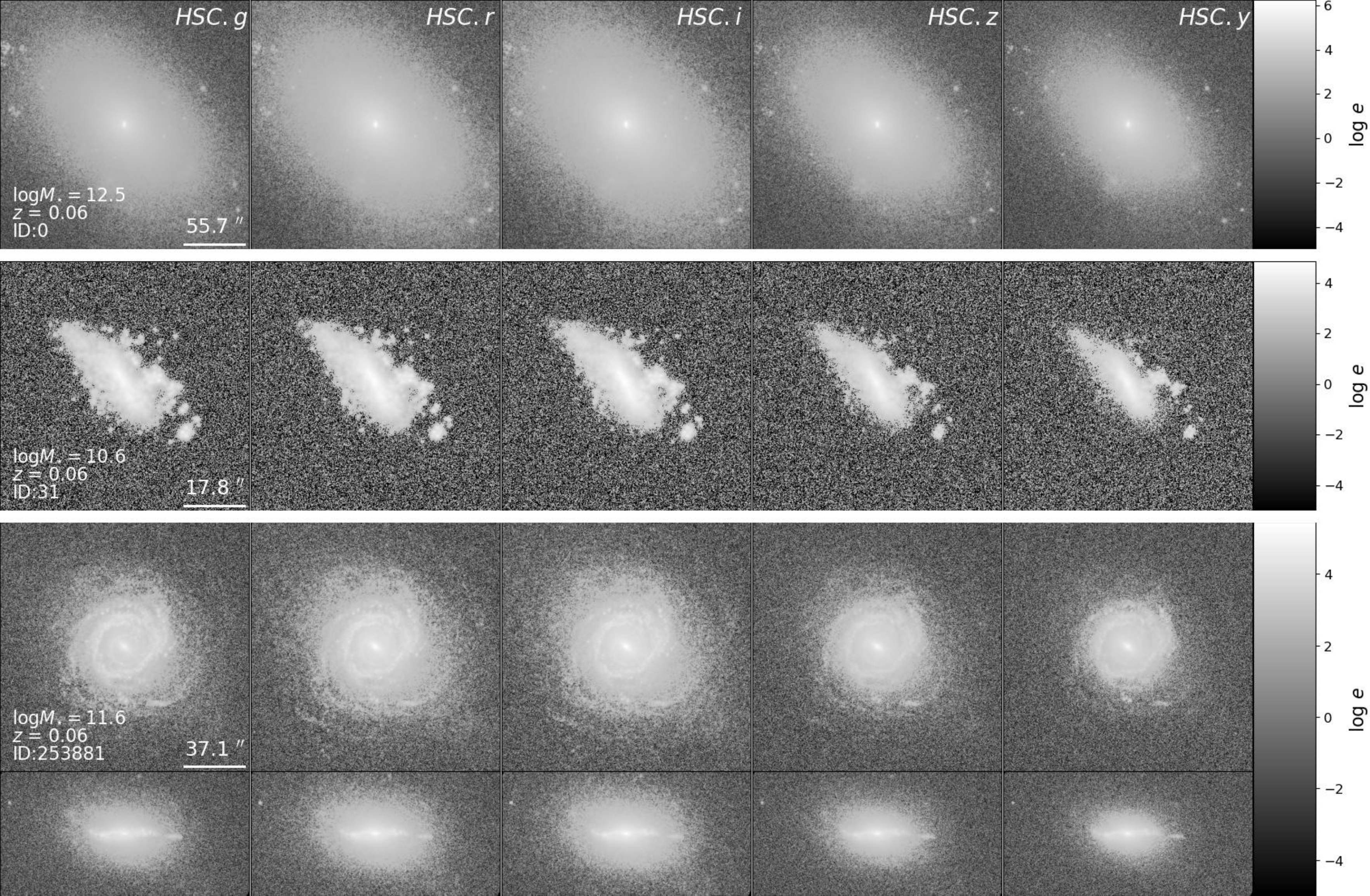}
    \caption{Mock galaxy images in $g,r,i,z$, and $y$ bands for HSC.}
    \label{fig: hsc}
\end{figure*}

The PSF describes how a point source of light, such as a star, is distributed across a detector, and is a fundamental characteristic of imaging system, as it determines the system's ability to resolve the fine details. The PSF size is influenced by fundamental physical limits as well as practical effects such as atmospheric turbulence, optical aberrations, and detector effects. In ground-based observations, the most dominant factor is atmospheric turbulence, while in space-based ones, detector effects play a larger role. Furthermore, the PSF exhibits variations across the position of the CCD and observational pointing. Consequently, for simplicity, in our galaxy generation for specific instrument, we only considered the PSF kernels at an arbitrary position and convolved the ideal images with these kernels. {We note that when convolving with the model PSF, the pixel sizes of the image and the kernel should be the same.}

{The addition of instrumental and sky background noises to a mock image simulates the various physical processes and external factors that can degrade the quality of the image signals. The instrumental noises involves several key components typically encountered in CCD imaging systems. The first category is shot noise, referred to the random fluctuation in the detected electrons by CCD due to the discrete nature of photons. This noise is modeled using a Poisson distribution with a mean equal to the expected electron count for each pixel as follows:}
\begin{equation}
    \hat{e}_{i, j}\ \sim\ {\rm Poisson}(\overline{e}_{i, j}),
\end{equation}
where $\hat{e}_{i, j}$ indicates the $i, j$ pixel of image on which the Poisson components are added; then, $\overline{e}_{i, j}$ is the mean of expected electron counts in $i, j$ pixel, calculated as:
\begin{equation}
    \bar{e}_{i, j} = e_{i, j} + N_{\rm mean},
\end{equation}
where $e_{i, j}$ is the pixels of image with PSF effects already applied; then,{ $N_{\rm mean}$ is the noise term that includes the contribution of two components, sky background noise, and dark current noise,}
\begin{equation}\label{eq: mean noise}
    N_{\rm mean} = (B_{\rm sky} + B_{\rm dark})t_{\rm exp}N_{\rm exp},
\end{equation}
{where $B_{\rm sky}$ and $B_{\rm dark}$ are sky background and dark current both in unit of electron count rate ($\rm e\ /\ s$). Between them, dark current, $B_{\rm dark}$, comes from thermal activity within CCD, even in the absence of light, while the sky background $B_{\rm sky}$ can be calculated by (see Section 9.4 in~\citet{Ryon2023}):}
\begin{equation}\label{eq: sky bkg}
    B_{\rm sky} = \frac{A_{\rm aper}{l_p^2}}{hc}\int\lambda I_{\rm sky}(\lambda)T(\lambda)d\lambda,
\end{equation}
{where $I_{\rm sky}$ is the brightness of the sky background in $\rm erg\ s^{-1}\ cm^{2}\ arcsec^2$ {\AA}$^{-1}$. The sky brightness is contributed by several components, such as airglow, zodiacal light, light pollution, and terrestrial background~\citep{Ryon2023}.} These components can vary depending on factors like calendar day, observational pointing, orbit or site position, and atmospheric conditions. Therefore, we only considered brightness curves under arbitrary conditions for specific observational campaigns. 

{Subsequently, readout noise was introduced, which results from the random fluctuations during the process of reading out the charge from a CCD.} This noise was modeled as Gaussian distribution with a mean of zero and a standard deviation corresponding to the readout level, $\sigma_{\rm RN}$, and is correlated with the number of exposures, $N_{\rm exp}$,
\begin{equation}
    \widetilde{e}_{i, j} \sim \hat{e}_{i, j} + N_{\rm exp}{\rm Gaussian(0, \sigma_{\rm RN})},
\end{equation}
where $\widetilde{e}_{i, j}$ represents the $i, j$ pixel of image with the readout noise added. Next, we subtracted the mean level of sky background and dark current from Equation~\ref{eq: mean noise} and expressed as
\begin{equation}
    e^{\rm obs}_{i, j} = \widetilde{e}_{i, j} - N_{\rm mean}.
\end{equation}
Finally, we obtained the galaxy images would be expected to be observed by specific instrument. The image unit in electron counts can be converted back to flux per frequency in $\rm Jy$. This procedure of including both the instrumental and sky background contributions ensures that the simulated data accurately represents the noisy conditions encountered in real-world astronomical observations. {Alternatively, the overall noise level for the galaxy images can also be determined from the limiting magnitudes for a specific band in a given survey~\citep{Fortuni2023, Bottrell2019, Bottrell2024, Martin2022, Wilkinson2024, Merlin2023}. This approach is also implemented and detailed in Appendix~\ref{app: derivation}. Depending on the availability of certain survey parameters, either this approach or the one described above can be utilized.}

\section{Sample outputs}\label{sec: sample outputs}
In this section, we showcase the generation capabilities of \texttt{GalaxyGenius} for a diverse range of photometric surveys conducted both in space and on ground. These surveys encompass notable ongoing and planned missions such as CSST, HST, JWST, Euclid, Roman, and HSC. Subsequently, we highlight the galaxies generated from EAGLE simulations. 
\subsection{CSST}\label{sec: CSST}
{The Chinese Space Station Telescope (CSST, ~\citet{Zhan2018, Gong2019}) is a 2-meter aperture space telescope}, planned to be launched at 2027. {This telescope is equiped with cutting-edge instruments,} including a spectrograph and a photometric system, enabling multi-wavelength observations and spectroscopic studies. The photometric system loads seven filters, $NUV, u, g, r, i, z$, and $y$ covering the near-UV (NUV) to NIR wavelength range. The pixel scale for each filter is uniform of $0.^{\prime\prime}074$. 

{Given that the filters of CSST barely reach NIR wavelengths, we utilized the ExtinctionOnly simulation framework to account for the extinction and scattering effects caused by dust components. Furthermore, we also employed the NoMedium mode for comparative purposes.} The wavelength range in rest-frame was set to be from 0.1 to 1.2 micron to account for all filters. Then the wavelength grid was configured as linear with 300 bins. {The exposure time, $t_{\rm exp}$, is assumed to be 150 s, with number of exposure, $N_{\rm exp}$, as four for the $NUV$ and $y$ bands and two for the other bands, according to the focal plane design.} {The sky backgrounds $B_{\rm sky}$ are calculated as $0.0042,0.0317,0.2803,0.3591,0.3708,0.2168$, and $0.0638\ \rm e\ / \ s$ based on the high sky background emission} including earthshine and zodiacal light in optical wavelength coverage at earth orbit provided by HST~\footnote{\url{https://hst-docs.stsci.edu/acsihb/chapter-9-exposure-time-calculations/9-7-tabular-sky-backgrounds}}. {Other necessary instrumental parameters employed in mock observation are outlined in Appendix~\ref{sec: instrumental parameters}.} We set the viewing angle, inclination and azimuth, to be both 0, meaning that the line of sight is in parallel to the $z$-axis. Additionally, for spiral galaxy, the viewing angles are set to be face-on and edge-on. These angles are calculated based on the total angular momentum for star particles within 30 kpc sphere~\citep{McAlpine2016}. 

Three subhalos are presented as examples and their properties are tabulated in Table~\ref{tab: metadata}, including stellar mass, star formation rate (SFR), and the number of distinct particles, as discussed in Section~\ref{sec: preprocess}. {The ideal SEDs generated from SKIRT in redshifted wavelength range are illustrated in Figure~\ref{fig: seds}. {The optical depth at 5500\text{\AA} $\tau_{5500\text{\AA}}$ is also calculated. For better comparison, deviations between different simulation modes are also displayed in the lower panel. The galaxy images in seven bands output by postprocess procedure introduced in Section~\ref{sec: postprocess} are depicted in Figure~\ref{fig: csst}.} The images are presented in logarithmic scale, and to mitigate the occurence of NaN values and enhance visual clarity, they are clipped at $10^{-5}$. 

Subhalo 0 is an elliptical galaxy with a low SFR and a substantial number of BC03 particles. {The slight deviation in the lower panel of SED for subhalo 0 (left panel of Figure~\ref{fig: seds}) indicates that the low fraction of dust elements shown in Table~\ref{tab: metadata} has a negligible impact on the SEDs in optical,  irrespective of the simulation mode utilized, with the corresponding optical depth being $9.76\times10^{-4}$.} In contrast, subhalo 31 is an irregular galaxy characterized by a high SFR and a large number of particles belonging to MAPPINGS-III SED family, which are indicative of star-forming regions. {The median panel of Figure~\ref{fig: seds} exhibits low continuum and several pronounced emission lines,} such as H$\alpha$ 6563, [SIII]$\lambda\lambda\ 9069,9532,$ and Pa$\gamma$ 10941 from low to high wavelength. {Furthermore, the dust exhibit substantial impacts on the continuum and emission lines, leading to substantial deviations in flux amplitude in optical wavelengths and yielding a high optical depth of 0.36.} Subhalo 253881, on the other hand, is a spiral galaxy with a high SFR characterized by several of aforementioned emission lines. The continuum exhibits similar shape to subhalo 0, primarily due to the dominance of BC03 particles. {As anticipated, the SED exhibits a significant reduction in the continuum for both face-on and edge-on directions, with attenuation more pronounced in the edge-on case. This is due to the increased dust column density encountered along the edge-on line of sight, resulting in an optical depth of 0.12 compared to 0.02 for the face-on view. These distinct behaviors demonstrate that the dust elements primarily attenuates emissions from star-forming regions, with the degree of attenuation increasing with the column density of dust through which the light propagates\citep{Lu2022}.}

\subsection{Euclid}\label{sec: Euclid}
The Euclid mission, developed by the European Space Agency (ESA), is a 1.2-meter space-based observatory designed to explore the nature of dark energy and dark matter by mapping the geometry of the Universe~\citep{Mellier2024}. Lauched in 2023, Euclid operates at the second Lagrange point (L2) and is optimized for both visible and NIR observations.

Euclid's payload includes two main instruments: VISible Imaging Channel (VIS, ~\citet{Cropper2024}) and Near Infrared Spectrometer and Photometer (NISP, ~\citet{Jahnke2024}). VIS covers the optical wavelength range from 500 to 900 nanometers with a pixel scale of $0.^{\prime\prime}1$, providing high-resolution, wide-field imaging. This instrument is designed for precise measurement of galaxy shapes, essential for weak gravitational lensing studies. While NISP operates in the NIR range of 920 to 2000 nm, including a photometric mode for imaging in three broad filters (Y, J, H) with pixel scales of $0.^{\prime\prime}3$, and a spectroscopic mode using a slitless grism. The spectroscopic mode is employed to measure redshifts of galaxies, rendering baryon acoustic oscillation (BAO) studies. 

We employed similar parameters to CSST (as mentioned in Section~\ref{sec: CSST}) to perform the simulation, and to cover the optical and NIR bands, the wavelength range are set to be 0.5 to 5 micron. Furthermore, DustEmission mode is activated to account for the radiative transfer process in IR bands. {And for comparative purposes, ExtinctionOnly mode is also performed and analyzed.} {We employed dust recipe as~\citet{Camps2016} (Equation~\ref{eq: camps}) and dust model as~\citet{Zubko2004} described in Section~\ref{sec: preprocess}.} The exposure time, $t_{\rm exp}$, are configured as 600 s with one exposure. {The PSFs are modeled using a Gaussian distribution with full width at half maximum (FHWM) as $0.204, 0.493, 0.515,$ and $0.553\ \rm arcsec$ for VIS, Y, J, and H bands respectively~\citep{Cropper2024, Jahnke2024}.} {The sky backgrounds are calculated as 0.7162, 3.6863, 4.4556, and 3.8228 $\rm e\ /\ s$ for these bands based on background models~\footnote{\url{https://irsa.ipac.caltech.edu/applications/BackgroundModel/}} considering that the telescope orbits at L2 point.  Other necessary instrumental parameters are given in Appendix~\ref{sec: instrumental parameters}. }

The SEDs in the NIR are presented in Figure~\ref{fig: seds}. {We notice that the NIR SEDs closely follow those in the optical wavelengths, indicating that extinction and scattering are the dominant effects in both optical and NIR wavelength ranges. Additionally, the deviations between DustEmission and ExtinctionOnly are negligible, except for several emission lines at longer wavelength. This suggest that dust emission has a minimal impact within the IR wavelength range ($0.5\mu {\rm m}\sim5\mu {\rm m}$) considered in this study.} Correspondingly, the galaxy images for the VIS and NISP bands are depicted in Figure~\ref{fig: euclid}. The NIR images reveal extended structures in comparison to the observations at optical bands.

\subsection{HST WFC3}\label{sec: HST WFC3}
The Hubble Space Telescope (HST; ~\citet{Freedman2001, Momcheva2016}) is a 2.4-meter optical and UV space observatory launched in 1990, orbiting Earth at an altitude of approximately 540 km. HST operates across UV (115-320 nm), visible (320-760 nm), and NIR (760-2500 nm) wavelength ranges, providing high-resolution imaging and spectroscopy free from atmospheric distortion. The telescope's pointing precision, fine guidance sensors, and advanced suite of instruments enable observations of faint and distance objects with unprecedented clarity.

The Wide Field Camera 3 (WFC3, ~\citet{Momcheva2016,Windhorst2011,Oconnor2024}), installed during the fifth servicing mission in 2009, is one of HST's most versatile instruments. It features two independent optical channels: UVIS and IR channel. UVIS operates in 200-1000 nm wavelength range with a $4096\times4096$ pixel CCD detector, optimized for UV and visible light imaging. This channel has a pixel scale of $0.^{\prime\prime}04$ and a field of view of approximately $162\times162\ \rm arcsec$; meanwhile the IR channel covers 800-1700 nm with a $1024\times1024$ HgCdTe array detector, with a pixel scale of $0.^{\prime\prime}13$, providing a field of view of $136\times123\ \rm arcsec$. For more detailed information on instrumental design, we refer to the WFC3 official website~\footnote{\url{https://www.stsci.edu/hst/instrumentation/wfc3/instrument-design/}}. 

The PSF for each filter is taken from the standard PSFs described in~\citep{Anderson2016}. The instrumental and sky background noises are calculated using the instrumental parameters and taking into account the operation of HST at Earth orbit. As an example, we considered several bands F275W, F390W, and F814W from UVIS channels and F105W, F110W, F125W, and F160W from the IR channels to illustrate the generation of galaxies from the TNG. The exposure time is similarly assumed to be 600 s with one exposure. The galaxy images in considered bands are depicted in Figure~\ref{fig: hst}. 

\subsection{JWST NIRCam}\label{sec: JWST NIRCam}
The James Webb Space Telescope (JWST, ~\citet{Sabelhaus2004}) is a cutting-edge observatory designed to operate primarily in the IR wavelength range, covering 0.6 to 28 microns. With a 6.5-meter segmented primary mirror, JWST provides a collecting area over six times larger than the HST, enabling unprecedented sensitivity and resolution for observations in the IR. 

The Near Infrared Camera (NIRCam) is JWST's primary imaging instrument, optimized for the 0.6 to 5.0 micron wavelength range. It is designed to perform wide-field imaging, high-resolution imaging and wavefront sensing for the alignment of JWST's segmented mirror. NIRCam comprises two identical optical modules, each with a $2.2\times2.2\ \rm arcminute$ field of view, providing redundancy and expanded coverage. A dichroic beam splitter separates light into short-wavelength (0.6-2.3 microns) and long-wavelength (2.4-5.0 microns) channels, enabling simultaneous observations in both spectral ranges. Each channel is equipped with a filter wheel that houses a variety of broadband, medium-band, and narrowband filter for photometry, as well as grisms for slitless spectroscopy. The pixel scales are $0.^{\prime\prime}031$ and $0.^{\prime\prime}063$ for the short- and long-wavelength channels respectively, enabling exquisite spatial resolution. For more detailed instrumental designs, please refer to NIRCam official website~\footnote{\url{https://jwst-docs.stsci.edu/jwst-near-infrared-camera/nircam-instrumentation/nircam-detector-overview}}.

{The PSF for each filter can be retrieved from STPSF~\footnote{\url{https://stpsf.readthedocs.io/en/latest/}}~\citep{Perrin2015}.} The background emission for JWST has been extensively studied in~\citet{Rigby2023} and can be obtained from JWST Background Tool (JBT)~\footnote{\url{https://github.com/spacetelescope/jwst_backgrounds}}. Given a pointing and observational date, the background emission including ISM, Zodiacal light, stray light, and thermal emission can be calculated from 0.5 to 30 micron. We similarly took a 600 s exposure and derive the background and instrumental noise for each band of JWST. {We took F070W, F150W, F200W, and F182M from the short-wavelength and F356W and F444W from long-wavelength channels, respectively. We display bandpass images in Figure~\ref{fig: jwst}.} 

\subsection{Roman}\label{sec: Roman}
{The Nancy Grace Roman Space Telescope (Roman, ~\citet{Spergel2015}), formerly known as WFIRST (Wide Field Infrared Survey Telescope),} is a next-generation 2.36 m space-observatory developed by NASA. Scheduled for launch in the mid-2020s, Roman is designed to address key astrophysical questions, including the nature of dark energy and dark matter, the detection of exoplanets, and the formation and evolution of galaxies. 

The Wide Field Instrument (WFI) on Roman operates primarily in the IR wavelength range, covering 0.48 to 2.3 microns, employ seven photometric filters and a grism for slitless spectroscopy. The pixel scales of each photometric filter is $0.^{\prime\prime}11$. The exposure time is similarly considered to be 600 s. We took the maximum value of dark current and readout noise demonstrated in the Roman user documentation~\footnote{\url{https://roman-docs.stsci.edu/roman-instruments-home}} for the calculation of instrumental and background noises, while the PSFs were obtained from STPSF for Roman given the name of filter. We illustrate the mock images in F062, F087, F106, F129, F158, F184, and F213 bands in Roman, respectively, in Figure~\ref{fig: roman}.

\subsection{HSC}\label{sec: HSC}
The Hyper Suprime-Cam Subaru Strategic Program (HSC-SSP, ~\citet{Aihara2018}) is a premier imaging survey conducted using the Hyper Suprime-Cam (HSC) on the 8.2-meter Subaru Telescope in Hawaii. It leverages HSC's 1.5-degree field of view and exceptional resolution with a median seeing $\sim0.6$ arcseconds to achieve deep, wide-field imaging across three tiers: wide (1,400 square degrees, $i\sim26$), deep (26 square degrees, $i\sim27$), and ultra-deep (3.5 square degres, $i\sim28$). The survey employs multi-band photometry in five broad bands ($g, r, i, z$, and $y$) and specialized narrowband filters, enabling precise photometric redshift estimations and the detection of emission line galaxies. Built on the LSST software framework, the HSC-SSP pipeline handles large-scale data reduction, providing calibrated images, coadded mosaics, and comprehensive catalogs~\citep{Bosch2018}. Scientifically, HSC-SSP addresses a wide range of objectives, including mapping dark matter via weak lensing, studying galaxy evolution, and probing the early universe with high-redshift galaxy detections, making it a cornerstone for cosmological and astrophysical research. 

The PSFs for five bands are obtained from the third Public Data Release~\citet{Aihara2018} using the \texttt{unagi} project~\footnote{\url{https://github.com/dr-guangtou/unagi}}. The background and instrumental noises are calculated by considering the sky emission at site Maunakea~\footnote{\url{https://www.gemini.edu/observing/telescopes-and-sites/sites\#OptSky}} and assuming an exposure time of 600 s. The generated images are displayed in Figure~\ref{fig: hsc}. We observe that the images differ from those in the same bands of CSST in Figure~\ref{fig: csst}. This is because the PSFs are significantly different between ground-based and space-based observations.

\subsection{Pseudo-color images}\label{sec: rgb}
To visualize the sample outcomes of \texttt{GalaxyGenius}, we generated pseudo-color images of selected spiral galaxies for CSST, Euclid, HST, JWST, Roman, and HSC, using the \texttt{make\_rgb} routine in \texttt{Astropy}~\footnote{\url{https://docs.astropy.org/en/stable/visualization/rgb.html\#}} based on the data products described in the previous subsections. However, it is important to note that optimal parameters for creating pseudo-color images require an evaluation across a wide field for various observations, while we employed the default settings ~\footnote{\url{https://docs.astropy.org/en/stable/visualization/rgb.html\#rgb-images-using-arbitrary-stretching}} for the purpose of demonstration.

In the upper panel of Figure~\ref{fig: rgb}, {we present subhalo 253881 at $z=0.06$ (snapshot 94) from IllustrisTNG.} The photometric bands employed are ($g, r, z$) for CSST, (Y, J, H) for Euclid, (IR\_F105W, IR\_F125W, IR\_160W) for HST, (F070W, F150W, F200W) for JWST, (F087, F129, F184) for Roman, and ($g, r, z$) for HSC respectively. The sizes of the cutouts are $74.4^{\prime\prime}\times74.4^{\prime\prime}$ for the face-on case (upper row) and $74.4^{\prime\prime}\times37.2^{\prime\prime}$ for the edge-on case (lower row). As shown in panel (a), the fine structures of the galaxy are resolved to varying levels by different telescopes. However, the differences are insignificant, given that the selected galaxy is both nearby and relatively large.

To further illustrate the resolving power of the aforementioned instruments, {we selected another spiral galaxy with a subhaloID of 11 at $z=0.20$ (snapshot 84)}. The corresponding pseudo-color images can be created with similar processes to those for generating the images in the panel (a) in Figure~\ref{fig: rgb}, but the sizes of cutouts differ due to the angular size of the mock images, which are $19.2^{\prime\prime}\times19.2^{\prime\prime}$ for the face-on case (upper row) and $19.2^{\prime\prime}\times9.6^{\prime\prime}$ for the edge-on case (lower row), respectively. As displayed in the panel (b), the effects of PSF smearing are much more pronounced than the galaxy at $z = 0.06$, highlighting the significance of using \texttt{GalaxyGenius} for the investigation of galaxy morphologies in more distant galaxies observed with space-borne telescopes.

\begin{figure*}
    \centering
    \begin{subfigure}[b]{1\textwidth}
        \centering
        \includegraphics[width=\textwidth]{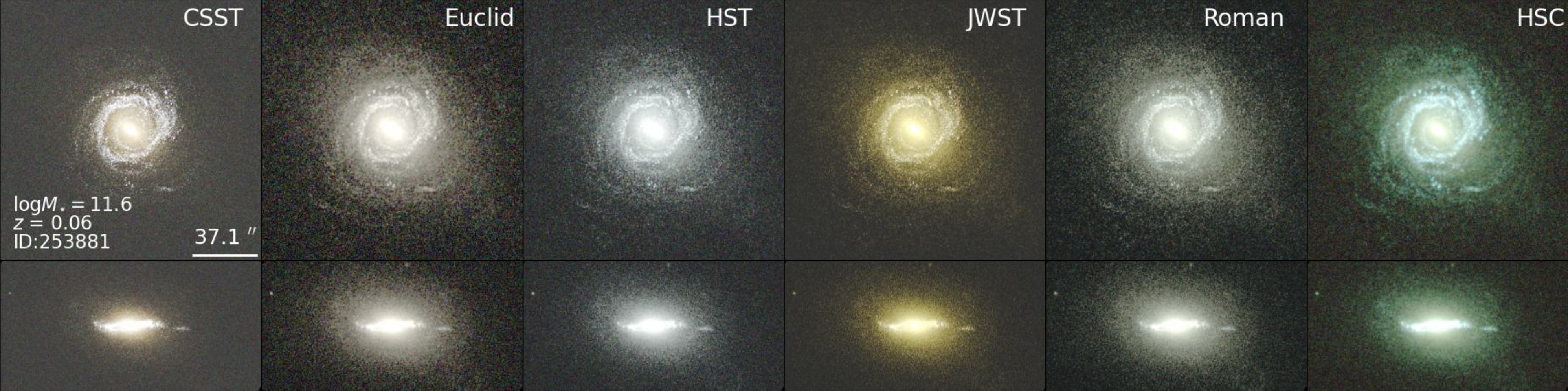}
        \caption{Subhalo 253881 in snapshot-94 at $z=0.06$}
    \end{subfigure}
    \\
    \begin{subfigure}[b]{1\textwidth}
        \centering
        \includegraphics[width=\textwidth]{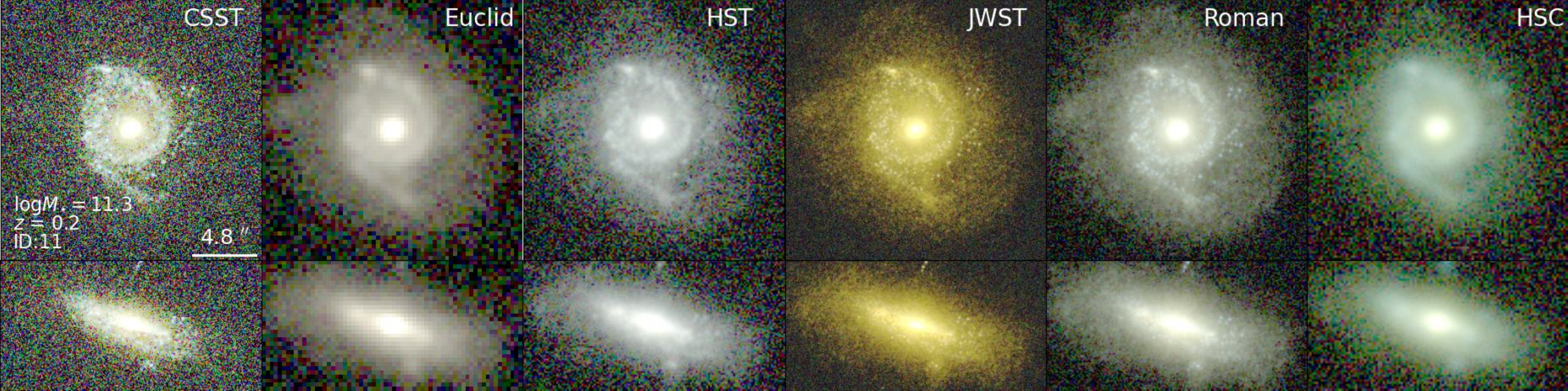}
        \caption{Subhalo 11 in snapshot-84 at $z=0.20$}
    \end{subfigure}
    \caption{Pseudo-color images constructed by \texttt{Astropy} utilizing default settings for CSST, Euclid, HST, JWST, Roman, and HSC for subhalo 253881 in snapshot-94 at $z=0.06$ discussed above (a) and subhalo 11 in snapshot-84 at $z = 0.20$ (b) respectively.}
    \label{fig: rgb}
\end{figure*}

\subsection{EAGLE simulation}\label{sec: eagle}

\begin{figure*}
    \centering
    \begin{subfigure}[b]{1\textwidth}
        \centering
        \includegraphics[width=\textwidth]{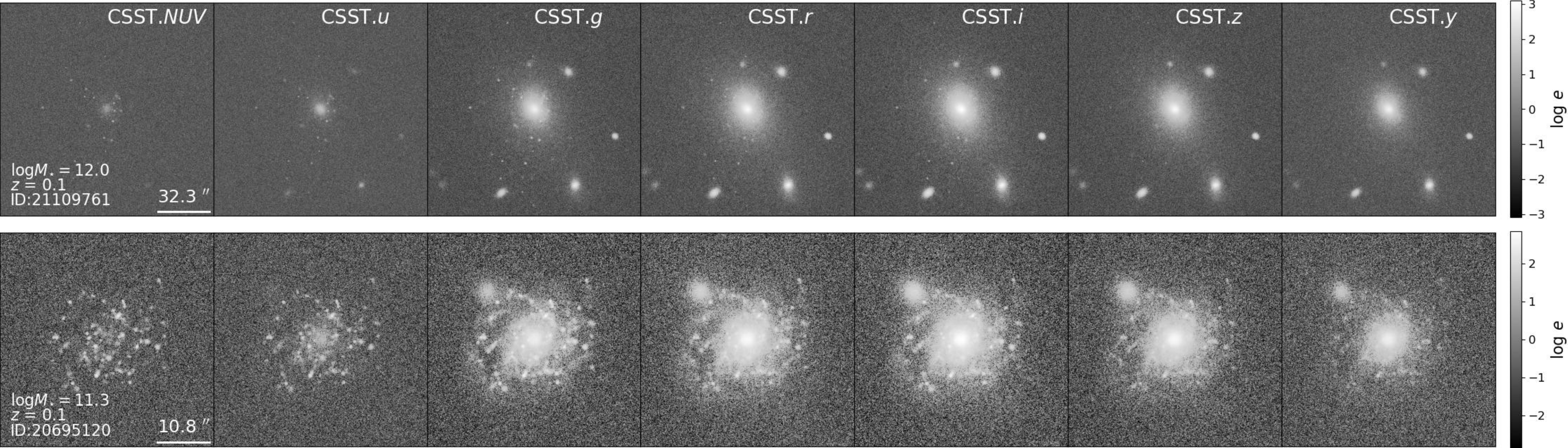}
        \caption{}
        \label{fig: csst eagle}
    \end{subfigure}

    \begin{subfigure}[b]{1\textwidth}
        \centering
        \includegraphics[width=\textwidth]{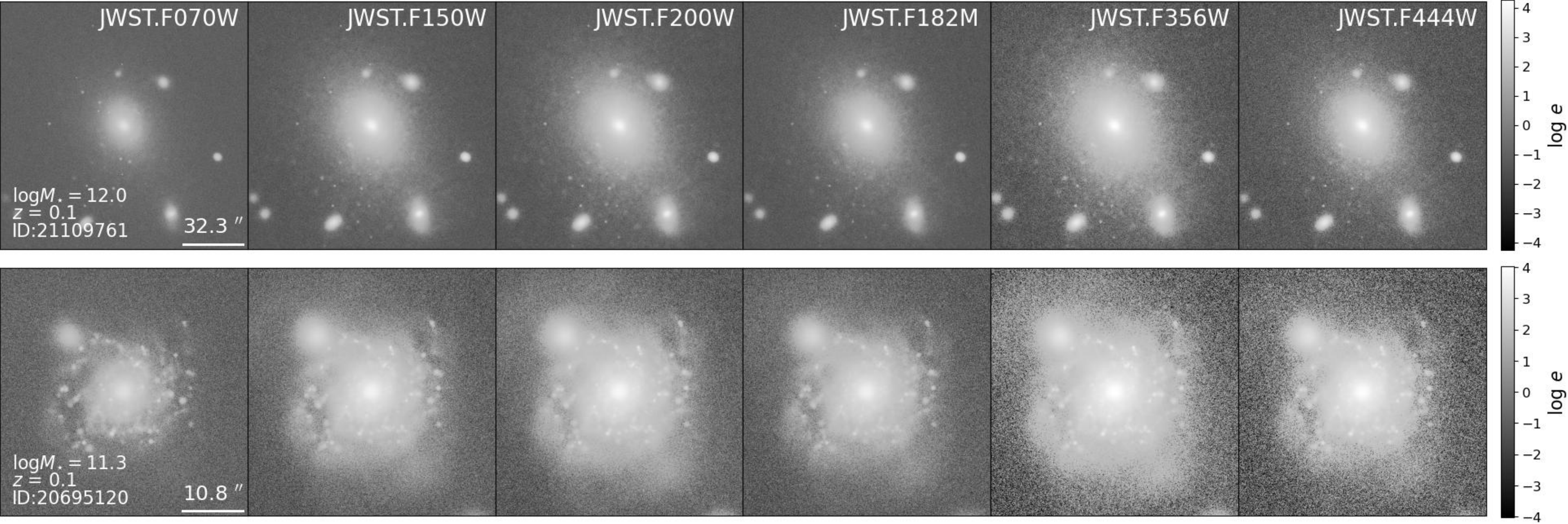}
        \caption{}
        \label{fig: jwst eagle}
    \end{subfigure}
    
    \caption{Mock images in $NUV, u, g, r, i, z$, and $y$ bands for CSST (a) and in F070W, F150W, F200W, F182M, F356W, and F444W for JWST (b) generated from EAGLE simulation using \texttt{GalaxyGenius}. {We notice that several companion sources exist apart from central targets. This is because that unlike TNG dataset, which stores its particle data based on individual subhalos, the particles retrieved from EAGLE are constraint by specific box length. Consequently, particles from companion sources are also included.}}
    \label{fig: eagle}
\end{figure*}

Galaxy images generated utill now are from IllustrisTNG simulations. However, \texttt{GalaxyGenius} is not constraint to specific hydrodynamical simulations. To demonstrate the capability of generating galaxies for general simulation, we employed EAGLE simulation as a case study. Two galaxies with GalaxyIDs of 21109761 and 20595120 at $z=0.101$ from RefL0100N1504 simulation in the EAGLE suite are considered. The parameters for these galaxies required by interface of data preprocessing module are acquired from SQL system in EAGLE's official website~\footnote{\url{https://icc.dur.ac.uk/Eagle/database.php}}. Stars, star-forming regions, and dust elements and their corresponding properties outlined in Table~\ref{tab: parameters} were retrieved from snapshot data using \texttt{pyread\_eagle}~\footnote{\url{https://github.com/kyleaoman/pyread_eagle}}. {It is important to note that unlike TNG simulations, the smoothing length for gas particles should also be extracted. This is because EAGLE simulations are typically conducted using a smoothed particle hydrodynamics (SPH) method known as Gadget-2~\citep{Springel2005} for both stars and gases.} 

Using these inputs, the preprocessing module creates the execution file for SKIRT. Once this step is complete, the subsequent procedures are identical to those used for IllustrisTNG simulations. Galaxy images anticipated by CSST and JWST are generated and displayed in Figure~\ref{fig: eagle}. We notice that they are elliptical and spiral galaxies, respectively, consistent with results from EAGLE. {In contrast to the sample images presented in Section~\ref{sec: sample outputs}, Figure~\ref{sec: eagle} exhibits several companion sources around the central image. This phenomenon arises from the fact that unlike the TNG dataset, which stores its particle data based on individual subhalos, the particles from EAGLE are retrieved in a box based on central coordinate of a specific subhalo and box length (as mentioned in Section~\ref{sec: preprocess}). Consequently, particles from companion sources are also included. However, the companion sources may have some issues due to incomplete extractions limited by box length.} Anyway, the above analyses highlight the \texttt{GalaxyGenius}'s adaptability to different simulations and its ability to produce consistent results across varied datasets.

\begin{figure*}
    \centering
    \includegraphics[width=1.\linewidth]{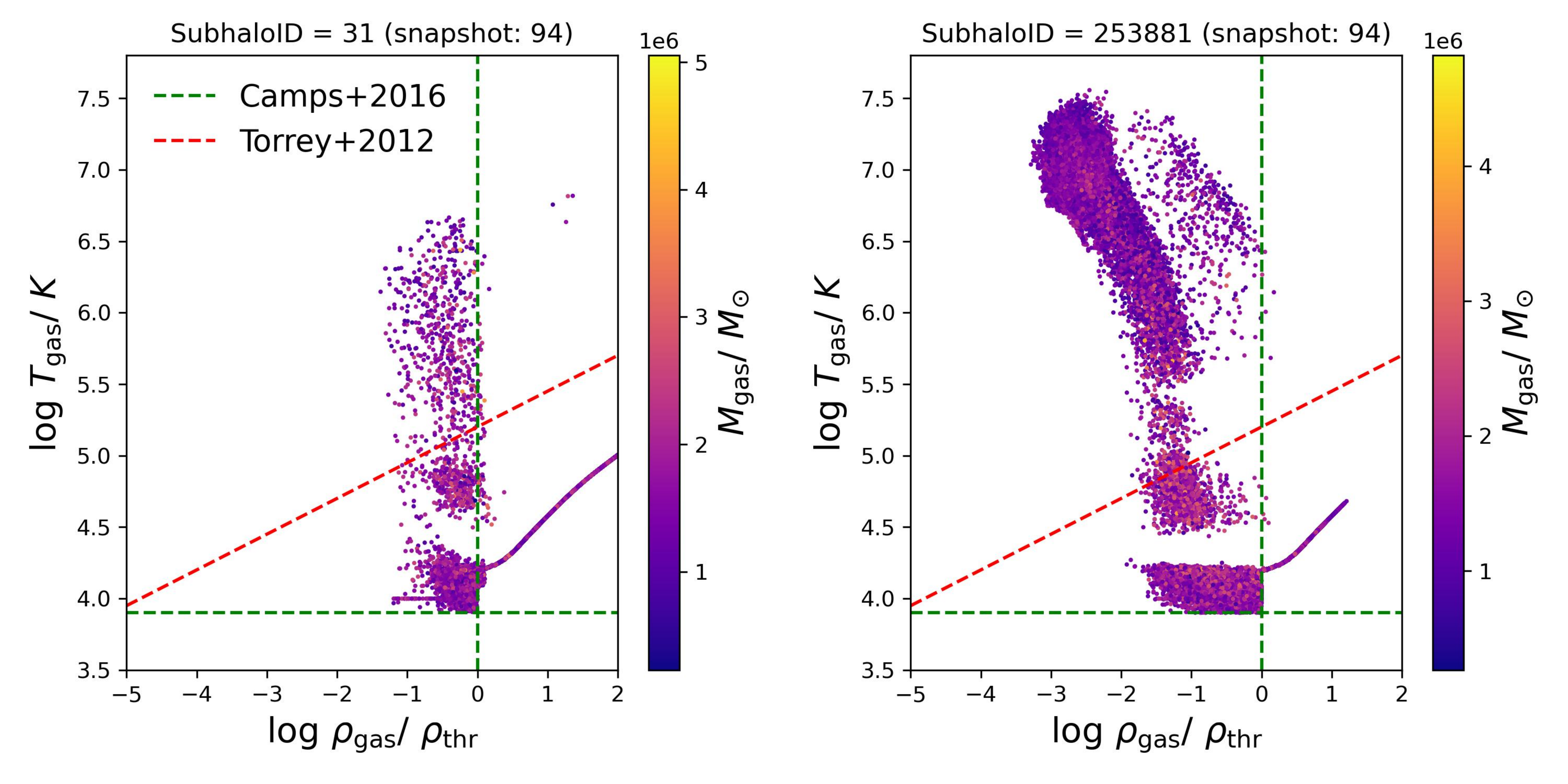}
    \caption{{Gas density versus temperature for subhalos 31 and 253881 using dust recipes as~\citet{Camps2016} (green dashed) and~\citet{Torrey2012} (red dashed) mentioned in Section~\ref{sec: preprocess}. The density is divided by a threshold distinguishing star-formation and non-star-formation, and the colorbar indicates the gas masses. }}
    \label{fig: ism recipe}
\end{figure*}

\begin{figure*}
    \centering
    \includegraphics[width=1.\linewidth]{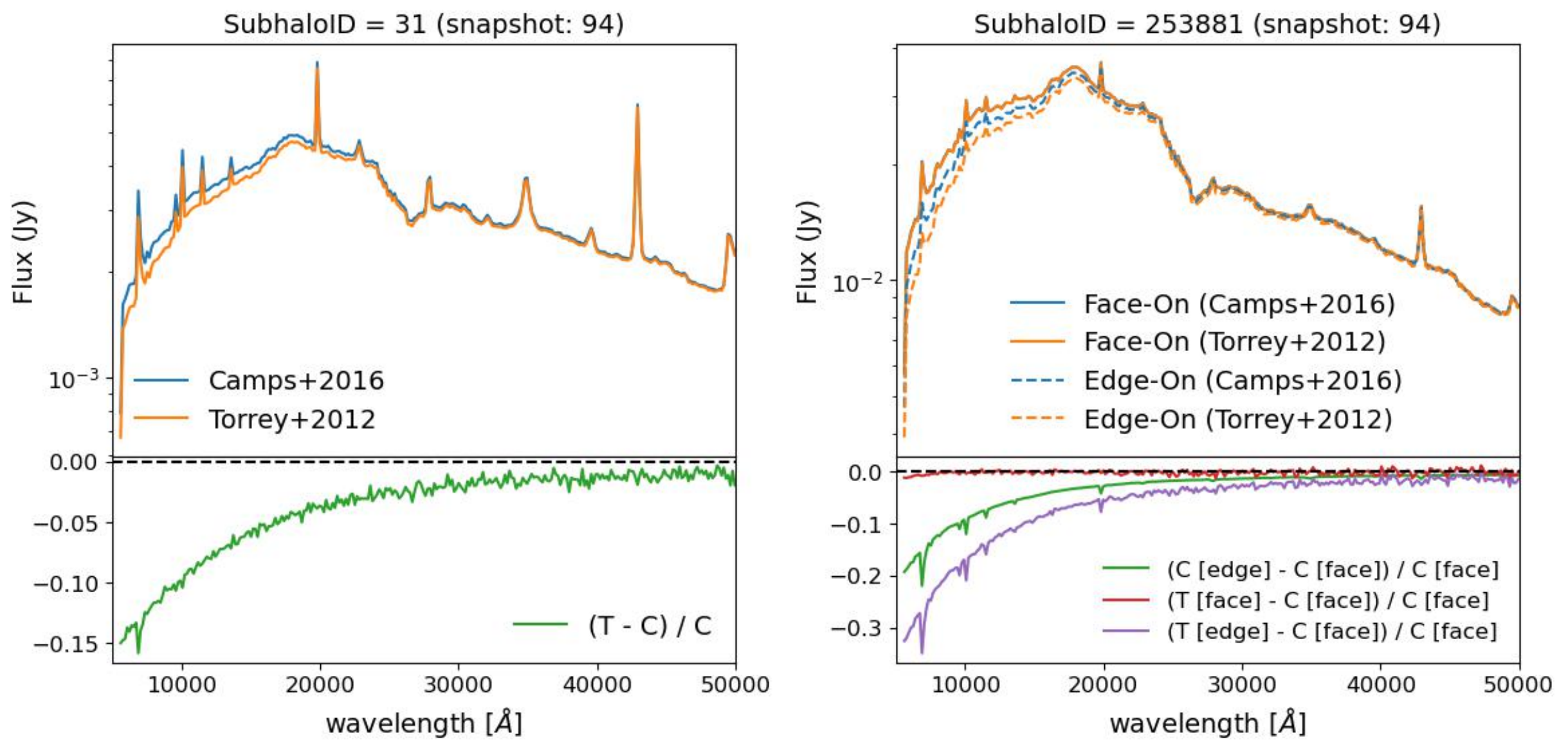}
    \caption{{Comparison of ideal SEDs generated by SKIRT in NIR for two dust recipes for subhalos 31 and 253881. The lower panels exhibit the normalized residual, where T and C indicate the dust recipes as~\citet{Torrey2012} and~\citet{Camps2016}}. }
    \label{fig: ism recipe seds}
\end{figure*}

\section{Discussion}\label{sec: discussion}
In this section, we first offer several discussions on dust recipes and dust models. Then we explain some limitations of generating galaxies from hydrodynamical simulations using the SKIRT project.

\subsection{Dust}\label{sec: dusts}
First, we investigated the impact of two dust recipes mentioned in Section~\ref{sec: preprocess} on galaxy generation. {Figure~\ref{fig: ism recipe} presents a scatter plot for gas density versus temperature within selected box of subhalo 31 and 253881. The colorbar indicates the gas mases.} The densities were directly retrieved from TNG and  normalized by a threshold that distinguishes between star formation and non-star formation. The temperatures are calculated using Equation~\ref{eq: temperature}. The two recipes~\citet{Camps2016} and~\citet{Torrey2012} are indicated by green and red dashed lines, respectively. 

We observe that the recipe proposed by~\citet{Torrey2012} tends to incorporate a higher number of dust particles. The precise numbers are presented in Table~\ref{tab: metadata}. {Ideal NIR SEDs generated by SKIRT for both subhalos are displayed in Figure~\ref{fig: ism recipe seds}, where the lower panels display the normalized residuals.} {From this figure, we notice that the fluxes of subhalo 31 exhibits a decrease in lower wavelength range when employing the recipe by~\citet{Torrey2012}.} {For subhalo 253881, the SEDs at face-on direction are identical regardless of the recipe, while edge-on SEDs all experience a decrease at similar wavelength range.} {Additionally, the flux decrease is more prominent for~\citet{Torrey2012} recipe, due to more dust introduced.} The observed phenomenon of subhalo 253881 at face-on direction can be potentially attributed to the substantial number of BC03 particles, leading to absence of any impact from additional dust on star-forming regions. Figure~\ref{fig: part dist} depicts the spatial distribution of dust and star-forming regions for both dust recipes, revealing that the recipe given by~\citet{Torrey2012} covers a larger area compared to the one given by~\citet{Camps2016}, which helps explain the SED behavior of subhalo 253881 observed in Figure~\ref{fig: ism recipe seds}. Additionally, as is shown in the right panel, for spiral galaxies such as subhalo 253881, the~\citet{Torrey2012} recipe provides a more complete dust distribution around the disk plane. 

On the other hand, we explore the influence of dust models that SKIRT project incorporated, including Zubko~\citep{Zubko2004}, DraineLi~\citep{Draine2007}, and Themis~\citep{Jones2017}. Themis comprises two families of dust particles: amorphous silicates and amorphous hydrocarbons, while Zubko and DraineLi also include polycyclic aromatic hydrocarbons (PAH) grains. The number of grain size bins employed to discretize the thermal emission calculations can be customized. In this study, we adopt 15 grain size bins for each dust mixture population, as recommended by~\citet{Kapoor2021}. The impact of employing distinct dust models on the SEDs is depicted in Figure~\ref{fig: dust model}. We notice that dust models indeed modify the flux curve for highly star-forming galaxies, which are predominantly composed of star-forming regions. {While for galaxies with limited star-formation such as subhalo 253881, this effect is negligible in considered wavelength range.}

{Additionally, this study only implement the dust attenuation effect from the dust particles inside galaxies themselves. A rigorous mock observation may also necessitate the dust extinction effect along the line-of-sight, i.e. the dust within the  IGM and the Milky Way~\citep{Thomas2017, Thomas2020, Thomas2021,Caredelli1989,Fitzpatrick1999,Sofue2016,Misiriotis2006}. However, it is generally believed that the dust extinction effect from the IGM is not significant, where the observational constraint is also very limited~\citep{Xie2015}. The foreground Milky Way dust extinction mainly depends on the galactic latitude of the target, which  has been extensively studied and is easily to calibrate~\citep{SFD, Li2018,Zhang2023,Zhang2025}. Users can implement their own modules to include and calibrate these effects. }

\begin{figure*}
    \centering
    \includegraphics[width=1.\textwidth]{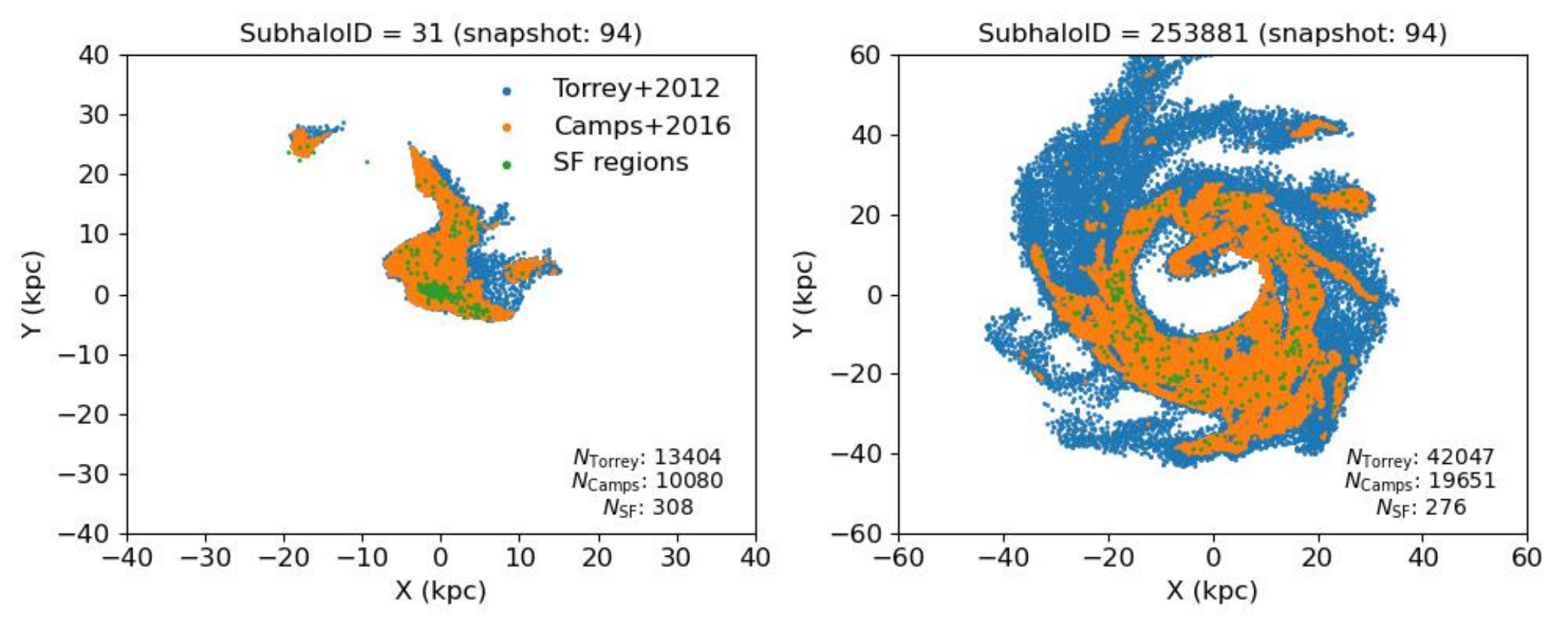}
    \caption{Spatial distributions on X-Y plane for star-forming regions and dust elements derived from two distinct recipes for subhalo 31 and 253881. {The number of star-forming regions and dust elements are also displayed in both plots.}}
    \label{fig: part dist}
\end{figure*}

\begin{figure*}
    \centering
    \includegraphics[width=1.\linewidth]{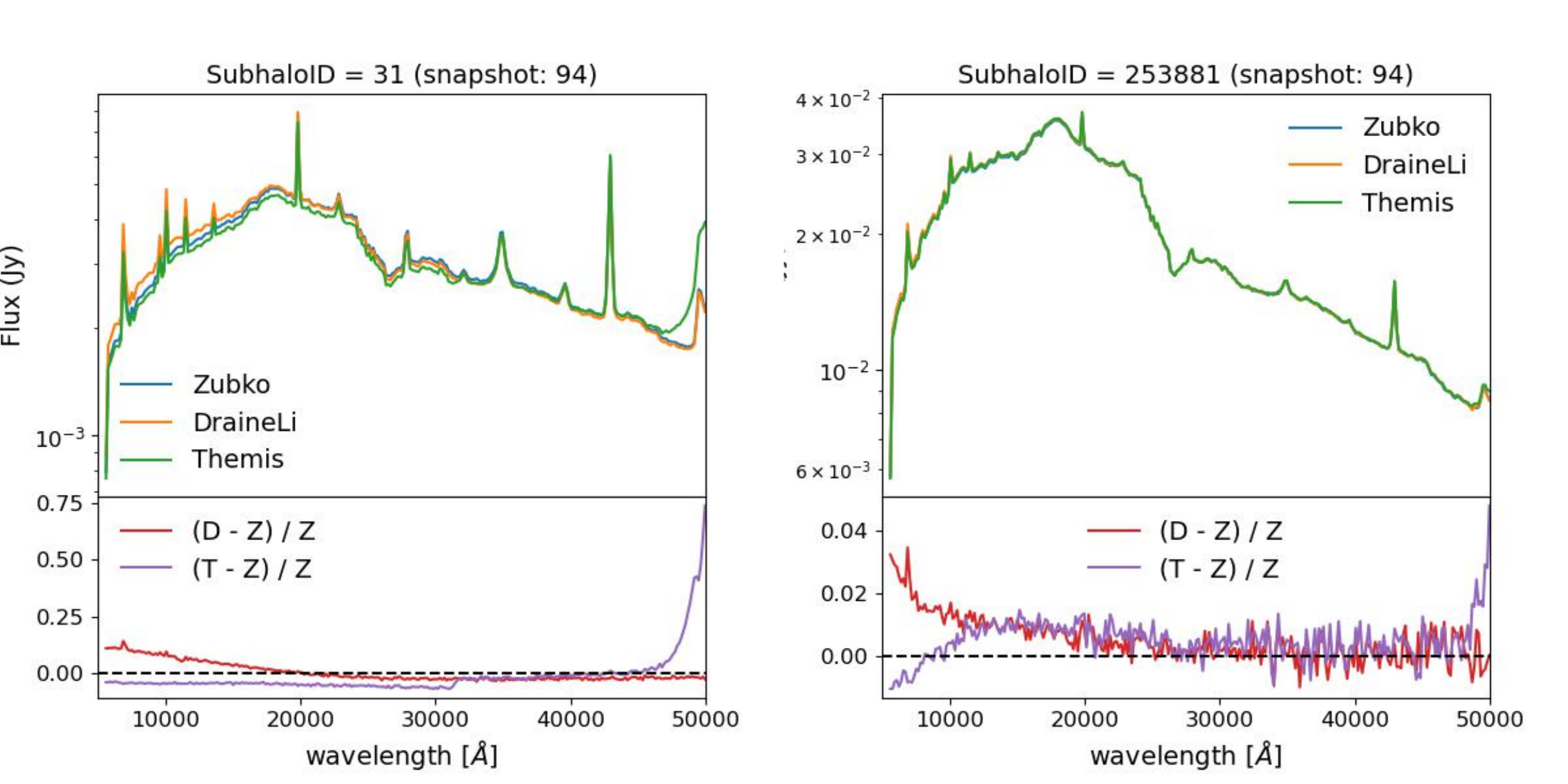}
    \caption{{Comparison of ideal SEDs generated using three dust models for subhalo 31 and 253881. The lower panels exhibit the normalized residual, where D, Z, and T indicate DraineLi~\citep{Draine2007}, Zubko~\citep{Zubko2004}, and Themis~\citep{Jones2017} dust models, respectively.}}
    \label{fig: dust model}
\end{figure*}

\subsection{Limitations}\label{sec: limitations}
{Current hydrodynamical simulations have significantly advanced our understanding of galaxy formation and evolution by modeling large cosmological volumes with high resolution.} However, they are subject to several limitations. One major constraint is spatial and mass resolution~\citep{Pillepich2018}, which affects the ability to resolve small-scale structure such as individual stars or molecular clouds. Limited resolution can lead to oversimplifed representations of physical processes, such as star formation and feedback mechanisms, which are often implemented via subgrid models, rather than being explicitly resolved. Additionally, the assumptions in subgrid physics, such as recipes for stellar feedback, black hole accretion, and gas cooling, can introduce uncertainties and biases in simulation outcomes~\citep{Matsumoto2023}. Another significant challenge is the limited volume of simulations, which restricts their ability to fully capture the large-scale structures of the universe. This limitations also results in a limited number of subhalos available for generating galaxies. For instance, in snapshot-94 of TNG100 (which we use as an example in Section~\ref{sec: preprocess}), there are approximately 6,000 galaxies with stellar masses exceeding $10^{10}\ M_\odot$. While one potential workaround is to increase the number of galaxies by considering multiple viewing angles, this approach does not address the lack of diversity among the galaxies. Galaxies generated from the same subhalo share identical intrinsic properties, such as stellar mass,  SFR, and other characteristics, limiting the ability to explore a broader range of galactic properties and behaviors.

On the other hand, SKIRT radiative transfer project is a powerful tool to simulate the interaction of radiation with dust and gas in astrophysical environments. Nonetheless, it has its own set of limitations. SKIRT simulations, especially those involving complex geometries or high-resolution grids, can be computationally demanding, requiring substantial memory and processing power, which can restrict the feasibility of highly detailed simulations. In regions with extremely high optical depths, accurately modeling radiative transfer becomes challenging~\citep{krieger2023, Camps2018failure, Baes2019optical}, as photon packets may undergo numerous interactions, leading to convergence issues and increased computational demands. As with all Monte Carlo methods, SKIRT simulations are subject to statistical noise, particularly in regions with low photon packet counts; mitigating this noise requires launching a large number of photon packets, which further increases computational load. {Moreover, the SED families within SKIRT, such as BC03, FSPS, and MAPPINGS-III are usually employed at low and intermediate redshifts. Their applicability to high redshift galaxies remains an open issue. Studies indicate that a single SSP model may not fully} capture the complexities of early-type galaxies at high redshifts~\citep{LopezCorredoira_2017}, potentially affecting the realism of the generated SEDs and galaxy images. Additionally, MAPPINGS-III requires three free parameters as shown in Table~\ref{tab: parameters} and their values need to be determined through calibrations with real observations. 

These limitations underscore the importance of ongoing advancements in finer and larger hydrodynamical simulations, and the development of stellar populations at high redshift. By addressing these challenges, the astrophysical community can further enhance the accuracy and predictive power of these tools, enabling deeper insights into the complex processes shaping galaxies across the universe, thereby improving the realism and fidelity for simulated galaxies from hydrodynamical simulations. 

{Furthermore, in this work, we are primarily focused on generating galaxy images. In addition to bandpass images, with data cubes and SEDs, it is also possible to simulate observations by IFU~\citep{Harborne2019,Wang2024,Sarmiento2023,Bottrell2022,Nanni2023,Cakir2024}, slit spectrograph~\citep{Fagioli2018,david2021}, and slitless spectrograph~\citep{Kummel2009, Neveu2024, Taylor2020, Brammer2012, Wen2024, Zhou2024} in a straightforward way by implementing external functions for post-processing. }

\section{Summary}\label{sec: conclusion}
To adequately prepare for current and upcoming galaxy surveys in data processing and scientific analysis, we propose a Python package, \texttt{GalaxyGenius}, designed to generate galaxy images for various photometric surveys. It consists of three main modules (see Section~\ref{sec: overall framework}): data preprocessing, ideal data cube generation, and mock observation. Specifically, as shown in Section~\ref{sec: preprocess}, the preprocessing module extracts properties of various particles from snapshots of hydrodynamical simulations and creates execution files for the following procedures. Subsequently, the data cube generation module assigns the particles with SEDs, according to their types and properties and then performs the MC radiative transfer process with SKIRT to generate IFU-like ideal data cubes (as illustrated in Section~\ref{sec: datacube gen}). Finally, the mock observation module (see Section~\ref{sec: postprocess}) constructs realistic galaxy images considering the throughputs, PSF {and noise levels of a particular survey.}

We present sample outputs of \texttt{GalaxyGenius} in Section~\ref{sec: sample outputs}, showcasing a suite of ongoing and planned photometric surveys conducted on both ground-based and space-based platforms, including CSST (Figure~\ref{fig: csst}), Euclid (Figure~\ref{fig: euclid}), HST (Figure~\ref{fig: hst}), JWST (Figure~\ref{fig: jwst}), Roman (Figure~\ref{fig: roman}), and HSC (Figure~\ref{fig: hsc}), covering wavelengths from UV to IR. The mock images are primarily based on the IllustrisTNG simulation. We also illustrate some mock images generated based on the EAGLE simulation (Figure~\ref{fig: eagle}), demonstrating the capability of \texttt{GalaxyGenius} to generate galaxy images for arbitrary simulations provided with necessary information for the preprocessing interfaces. 

The current implementation of \texttt{GalaxyGenius} has several limitations, which arise from two primary sources. The first stems from hydrodynamical simulations, including constraints such as limited mass resolution, assumptions in physical models, and restricted simulation volumes. The second source of limitations lies in the SKIRT radiative transfer code, particularly its significant memory and computational overhead, as well as the applicability of integrated SED families for high redshift universe. Despite these limitations, the modular architecture of \texttt{GalaxyGenius} offers significant flexibility. Since its modules are isolated and independent, users can replace or customize individual components with their own models or algorithms to address specific limitations. 

In summary, \texttt{GalaxyGenius} is a modular framework that bridges the gap between hydrodynamical simulations and real observations. It can be used to prepare mock galaxy images to verify data processing pipelines, validate scientific analysis software, and support deep learning-relevant investigations when real observations are insufficient. Additionally, since \texttt{GalaxyGenius} is an open-source project, users have the flexibility to replace existing codes in the modules with their own, enabling the exploration beyond imaging surveys to include transients, IFU, slit or slitless spectra, and more.

\section{Data Availability}
The source code and documentation for \texttt{GalaxyGenius} are publicly available online at~\url{https://github.com/xczhou-astro/galaxyGenius/}.

\begin{acknowledgements}
We appreciate helpful suggestions and comments from Maarten Baes, Qi Zeng, Andrea Gebek, Nick Andreads and members of SKIRT team. XCZ and NL acknowledge the support from the science research grants from the China Manned Space Project (No. CMS-CSST-2021-A01), the CAS Project for Young Scientists in Basic Research (No. YSBR-062), and the Ministry of Science and Technology of China (No. 2020SKA0110100). H.Y. acknowledges the support from the National Natural Science Foundation of China (Grant No. 11988101). Z.H. acknowledges support from the China Postdoctoral Science Foundation under Grant Number GZC20232990 and the National Natural Science Foundation of China (Grant No. 12403104). A.I. acknowledges the Postdoctoral Research Fund (R517L(N)) from Guangdong Technion Israel Institute of Technology and the Alliance of International Science Organizations (Grant No. ANSO-VF-2024-01). CW acknowledges the support from the China Manned Space Project (No. CMSCSST-2021-A03). SS thanks for research grants from  the National Key R\&D Program of China (No. 2022YFF0503402), the China Manned Space Project with No. CMS-CSST-2025-A07, the National Natural Science Foundation of China (No. 12141302), and Shanghai Academic/Technology Research Leader (22XD1404200).CMS-CSST-2025-A07, National Natural Science Foundation of China (No. 12141302) and Shanghai Academic/Technology Research Leader (22XD1404200).
\end{acknowledgements}

%

\bibliographystyle{aa}
\bibliography{aa}

\begin{appendix}

\onecolumn
\section{Instrumental Parameters}\label{sec: instrumental parameters}
Instrumental parameters employed to perform mock observations for selected filters illustrated in Section~\ref{sec: sample outputs} for CSST, Euclid, HST WFC3, JWST NIRCam, Roman, and HSC are outlined in Figure~\ref{fig: instrumental parameters}. $\lambda_{5\%}$ and $\lambda_{95\%}$ demonstrate the wavelength range of individual filter. The sky backgrounds, $B_{\rm dark}$ are calculated using Equation~\ref{eq: sky bkg} employing sky emission curves for each survey. $B_{\rm dark},\ \sigma_{\rm RN},\ D_{\rm aper}$ and $l_{\rm p}$ are dark current, readout, diameter of telescope and pixel scale of filters respectively. For CSST, we use tabular sky emission in~\url{https://hst-docs.stsci.edu/acsihb/chapter-9-exposure-time-calculations/9-7-tabular-sky-backgrounds}. The instrumental parameters are obtained from private communications. Note that the number of exposure, $N_{\rm exp}$, are determined by the focal plane design. While for Euclid, the emission curve is obtained from background model in~\url{https://irsa.ipac.caltech.edu/applications/BackgroundModel/} by specifying coordinates at $\rm (ra,\ dec)=(150.09,\ 2.21)$ and observing location at Earth-Sun L2 point. The instrumental parameters are obtained from~\url{https://sci.esa.int/web/euclid/-/euclid-vis-instrument} and~\url{https://sci.esa.int/web/euclid/-/euclid-nisp-instrument} for VIS and NISP respectively. For UVIS filters of HST WFC3, we employ the same emission curve as for CSST, while for IR filters, the emission curve is obtained from background model used for Euclid but with observing location changed to Earth orbit. The instrumental parameters for UVIS and IR are provided in~\url{https://www.stsci.edu/hst/instrumentation/wfc3/instrument-design/}. The curve for JWST is acquired from JWST backgrounds tool (JBT) in~\url{https://github.com/spacetelescope/jwst_backgrounds} using default parameters. And instrumental parameters are retrieved from~\url{https://jwst-docs.stsci.edu/jwst-near-infrared-camera/nircam-instrumentation/nircam-detector-overview/nircam-detector-performance#gsc.tab=0}. For Roman, same background model used by Euclid is employed. We use the most pessimistic values for these instrumental parameters provided in~\url{https://roman.ipac.caltech.edu/sims/Param_db.html}. While for ground-based HSC, we employ sky curve at Mauna Kea site given in~\url{https://www.gemini.edu/observing/telescopes-and-sites/sites#OptSky}. The intrumental parameters employed are given in~\citet{Aihara2018}.

\begin{figure}
    \centering
    \includegraphics[width=1.\linewidth]{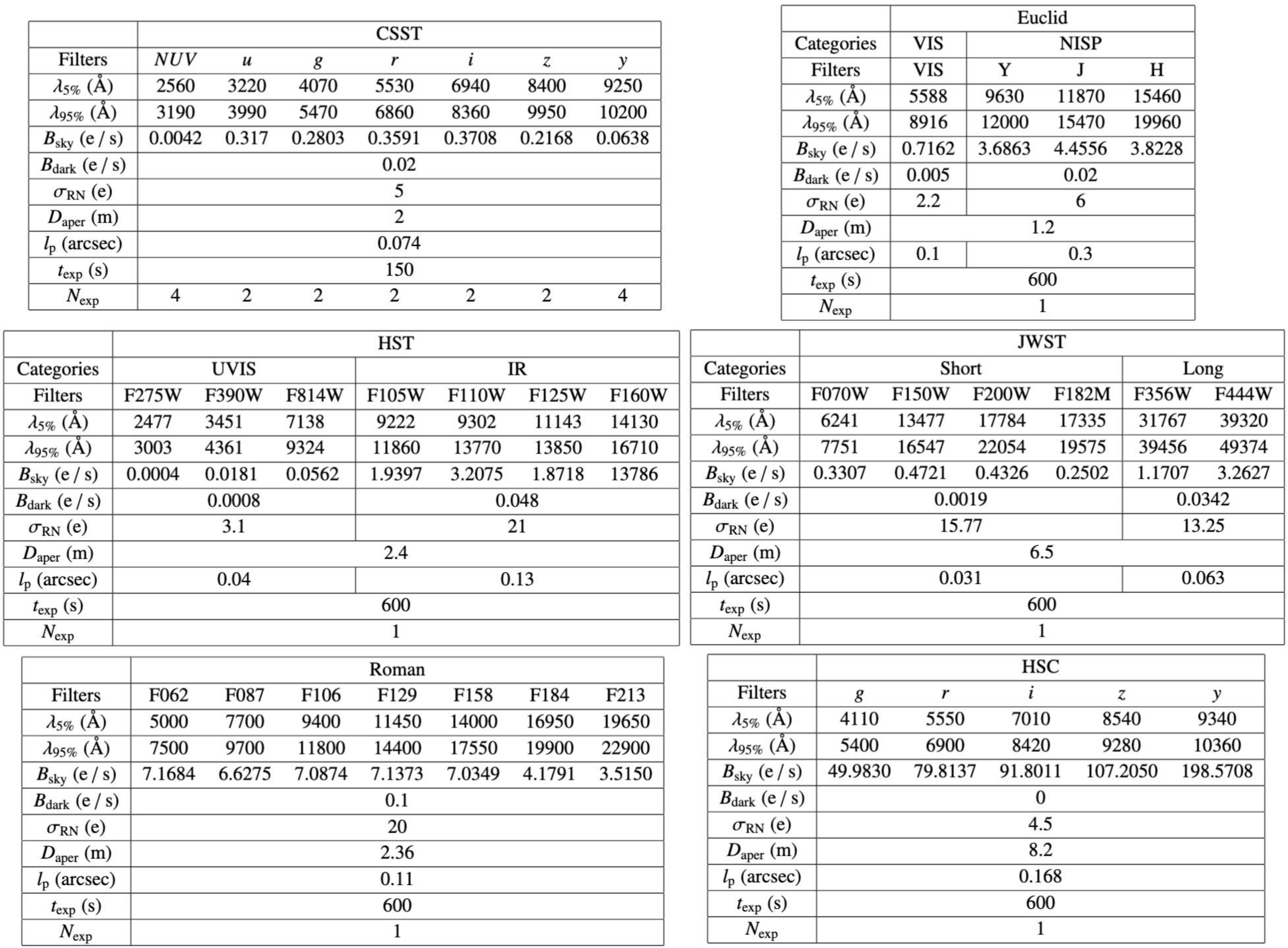}
    \caption{Instrumental parameters employed to perform mock observations for selected filters in Section~\ref{sec: sample outputs} for CSST, Euclid, HST WFC3, JWST NIRCam, Roman, and HSC.}
    \label{fig: instrumental parameters}
\end{figure}

\section{PSFs}\label{app: psfs}
The PSFs used to convolve with the ideal bandpass images in Section~\ref{sec: postprocess} are displayed in Figure~\ref{fig: psfs}. Each PSF is fixed at $5^{\prime\prime}$ and shown in log-scale with clipped value at 1e-5 for avoiding zero values and for enhanced illustration. For CSST, we obtain the PSFs by private communications. For Euclid, the PSFs are modeled using a Gaussian kernel {with FWHM as 0.204, 0.493, 0.515 and 0.553 $\rm arcsec$} for VIS, Y, J and H bands respectively~\citep{Cropper2024, Jahnke2024}. PSFs for HST are taken from standard PSFs described in~\citet{Anderson2016} and downloaded in~\url{https://www.stsci.edu/hst/instrumentation/wfc3/data-analysis/psf}. While for JWST and Roman, the PSFs are retrieved from STPSF in~\url{https://stpsf.readthedocs.io/en/latest/} given the filter names. Finally, HSC PSFs are obtained from the third Public Data Release(PDR3, ~\citet{Aihara2018}) using the \texttt{unagi} project in~\url{https://github.com/dr-guangtou/unagi} by specifying coordinate at $\rm (ra,\ dec)=(150.09,\ 2.21)$.

\begin{figure*}
    \begin{subfigure}[b]{1\textwidth}
        \centering
        \includegraphics[height=2cm]{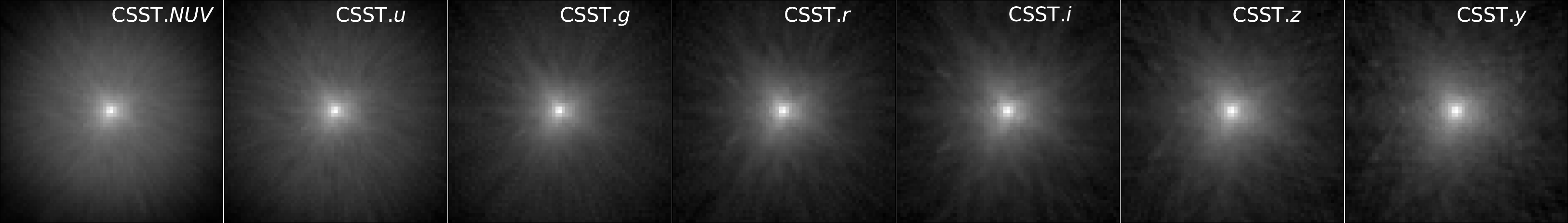}
    \end{subfigure}

    \begin{subfigure}[b]{1\textwidth}
        \centering
        \includegraphics[height=2cm]{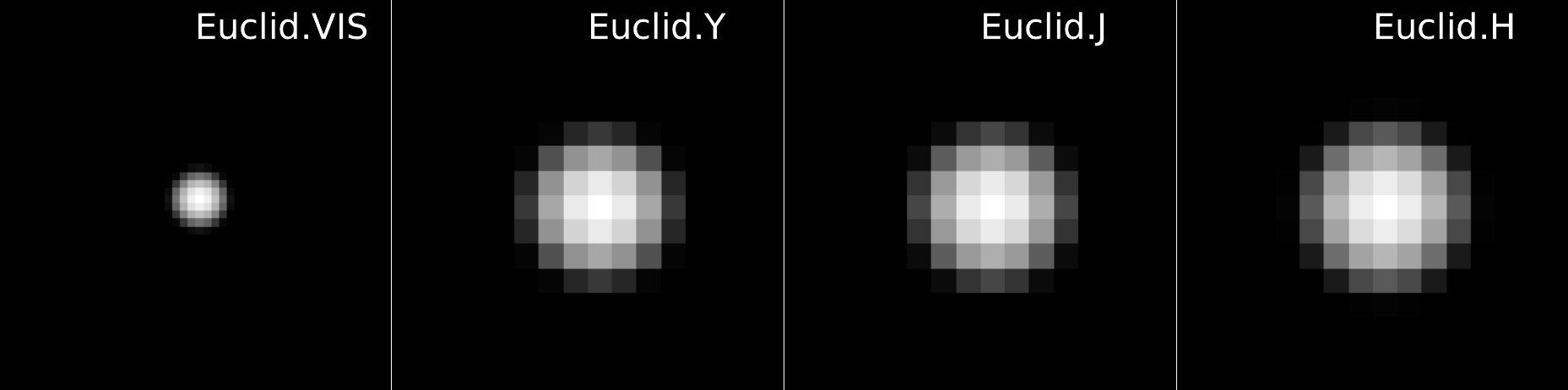}
    \end{subfigure}

    \begin{subfigure}[b]{1\textwidth}
        \centering
        \includegraphics[height=2cm]{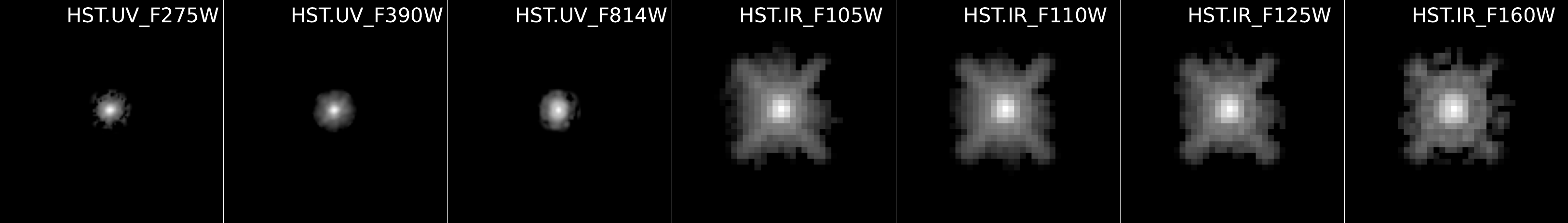}
    \end{subfigure}

    \begin{subfigure}[b]{1\textwidth}
        \centering
        \includegraphics[height=2cm]{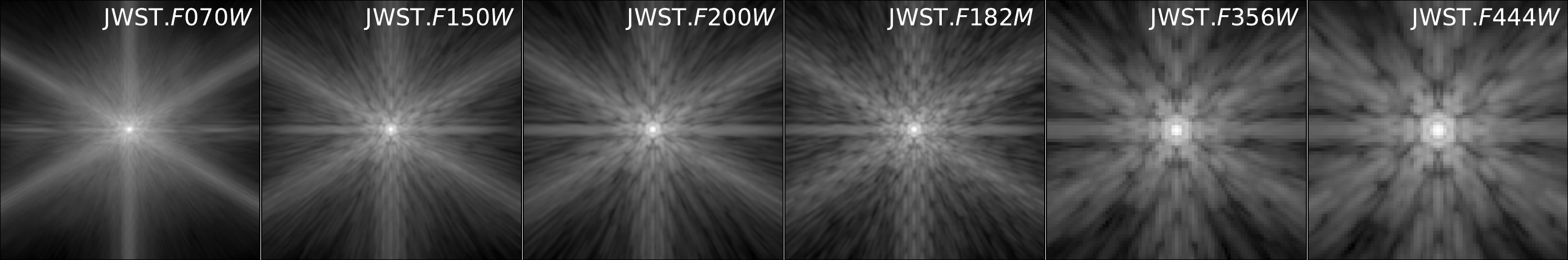}
    \end{subfigure}

    \begin{subfigure}[b]{1\textwidth}
        \centering
        \includegraphics[height=2cm]{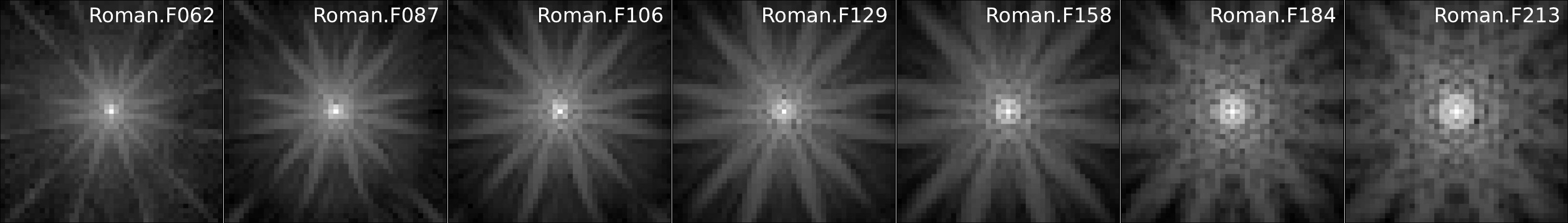}
    \end{subfigure}

    \begin{subfigure}[b]{1\textwidth}
        \centering
        \includegraphics[height=2cm]{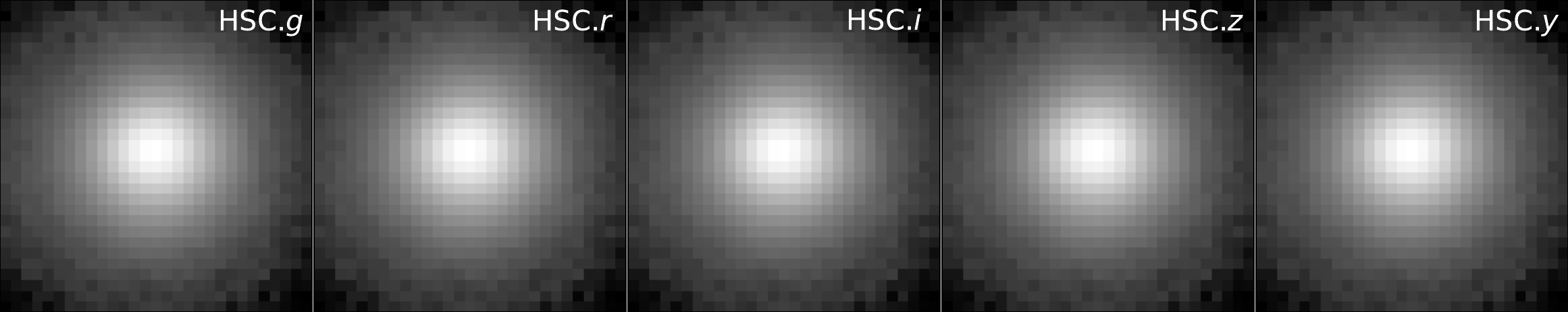}
    \end{subfigure}

    \caption{
Point spread functions (PSFs) are used to convolve with the ideal bandpass images presented in Section~\ref{sec: postprocess}. The size of each PSF is fixed at $5^{\prime\prime}$. Larger PSFs are cropped at the center, while smaller ones are padded with zeros. For enhanced visualization, these PSFs are displayed in a logarithmic scale, with clipped value of 1e-5 to avoid zero values. It is noteworthy that for the Euclid mission, PSFs are constructed as Gaussian models. }
    \label{fig: psfs}
\end{figure*}

\section{Noise level}\label{app: derivation}
{Assuming that a given limiting magnitude corresponds to the S/N $n_{\rm lim}$, we have
\begin{equation}\label{eq: snr}
    n_{\rm lim} = \frac{C_{\rm lim}}{\sqrt{C_{\rm lim}  + N_{\rm exp}t_{\rm exp}(B_{\rm sky} + B_{\rm dark})N_{\rm pix}^A + N_{\rm exp} \sigma_{\rm RN}^2N_{\rm pix}^A}}
,\end{equation}
where $C_{\rm lim}$ and $N_{\rm pix}^A$ are the electron counts and number of pixels in the aperture used to estimate the limiting magnitude. The electron counts $C_{\rm lim}$ can be calculated as:
\begin{equation}\label{eq: count}
\begin{split}
    C_{\rm lim} & = \frac{N_{\rm exp}t_{\rm exp}A_{\rm aper}}{hc}\int\lambda f_{\rm lim}(\lambda)T(\lambda)d\lambda \\
     & = \frac{N_{\rm exp}t_{\rm exp}A_{\rm aper}}{h}\frac{\int\frac{\lambda}{c} f_{\rm lim}(\lambda)T(\lambda)d\lambda}{\int{\frac{1}{\lambda} T(\lambda)d\lambda}}\int{\frac{1}{\lambda} T(\lambda)d\lambda}\\
     & = \frac{N_{\rm exp}t_{\rm exp}A_{\rm aper}}{h}\frac{\int\frac{1}{h\nu} f_{\rm lim}(\nu)T(\nu)d\nu}{\int{\frac{1}{h\nu} T(\nu)d\nu}}\int{\frac{1}{\lambda} T(\lambda)d\lambda}\\
     & = \frac{N_{\rm exp}t_{\rm exp}A_{\rm aper}}{h} 10^{\frac{{\rm ZP}-m_{\rm lim}}{2.5}}\int \frac{1}{\lambda} T(\lambda)d\lambda,
\end{split}
\end{equation}
where $f_{\rm lim}(\lambda)$, $\rm ZP$ and $m_{\rm lim}$ are flux in $\rm erg\ cm^{-2}\ s^{-1}\ $A$^{-1}$, zero-point and limiting magnitude respectively. We note that $f_{\rm lim}$ is already integrated over solid angle. $f_{\rm lim}(\nu)$ is in $\rm Jy$, converted from $f_{\rm lim}(\lambda)$ by Equation~\ref{eq: nu_to_lambda} with a pivot wavelength, $\lambda_p^2$, for a bandpass of
\begin{equation}
    \lambda_p = \sqrt{\frac{\int T(\lambda)d\lambda}{\int T(\lambda)\lambda^{-2}d\lambda}}
.\end{equation}
Here, we define gain factor $g$ as
\begin{equation}
    g = \frac{h}{A_{\rm aper}\int \frac{1}{\lambda} T(\lambda)d\lambda}
.\end{equation}
This factor is designed for conversion between electron counts and flux, and is only relevant to the observational instrument. Consequently, the electron counts can be written as
\begin{equation}\label{eq: C with g}
    C_{\rm lim} = \frac{N_{\rm exp}t_{\rm exp}}{g}10^{\frac{ZP - m_{\rm lim}}{2.5}}.
\end{equation}
Insert this equation to Equation~\ref{eq: snr} and then we have
\begin{equation}
\begin{split}
    N_{\rm exp}t_{\rm exp}(B_{\rm sky} + B_{\rm dark}) + N_{\rm exp}\sigma_{\rm RN}^2
    = \frac{1}{N_{\rm pix}^A}\left(\frac{C_{\rm lim}}{n_{\rm lim}}\right)^2 - \frac{C_{\rm lim}}{N_{\rm pix}^A}
    \\ = \left( \frac{N_{\rm exp}t_{\rm exp}}{gn_{\rm lim}} 10^{\frac{{\rm ZP}-m_{\rm lim}}{2.5}}\right)^2\frac{1}{N_{\rm pix}^A} - \frac{N_{\rm exp}t_{\rm exp}}{gN_{\rm pix}^A}10^{\frac{{\rm ZP}-m_{\rm lim}}{2.5}}
\end{split}
.\end{equation}
The noise level is the denominator of Equation~\ref{eq: snr}. Therefore, for a source with intensity, $I_{\rm src}$, in unit of $\rm Jy/sr$, the noise level, $\sigma_{\rm e}$, in electron counts would be:
\begin{equation}
\label{eq: sigma_e}
\begin{split}
    \sigma_{\rm e} = \sqrt{C + N_{\rm exp}t_{\exp}(B_{\rm sky} + B_{\rm dark}) + N_{\rm exp}\sigma_{\rm RN}^2}
    \\ = \sqrt{\frac{N_{\rm exp}t_{\rm exp}I_{\rm src}l_{\rm p}^2}{g} + \left( \frac{N_{\rm exp}t_{\rm exp}}{gn_{\rm lim}} 10^{\frac{{\rm ZP}-m_{\rm lim}}{2.5}}\right)^2\frac{1}{N_{\rm pix}^A} - \frac{N_{\rm exp}t_{\rm exp}}{gN_{\rm pix}^A}10^{\frac{{\rm ZP}-m_{\rm lim}}{2.5}}}
\end{split}
.\end{equation}
We note that the source intensity $I_{\rm src}$ in $\rm Jy/sr$ is converted to electron counts by gain factor $g$:
\begin{equation}
    \begin{split}
        C = \frac{N_{\rm exp}t_{\rm exp}I_{\rm src}l_{\rm p}^2}{g}
    \end{split}
.\end{equation}
Alternatively, the noise level can be expressed in $\rm Jy/sr$ via
\begin{equation}\label{eq: sigma}
\begin{split}
    \sigma & = \frac{g}{N_{\rm exp}t_{\rm exp}l_{\rm p}^2}\sigma_e 
    \\ & = \frac{1}{l_{\rm p}^2}\sqrt{\frac{gI_{\rm src}l_{\rm p}^2}{N_{\rm exp}t_{\rm exp}} + \left(\frac{1}{n_{\rm lim}}10^{\frac{{\rm ZP} - m_{\rm lim}}{2.5}}\right)^2\frac{1}{N_{\rm pix}^A} - \frac{g}{N_{\rm exp}t_{\rm exp}N_{\rm pix}^A}10^{\frac{{\rm ZP} - m_{\rm lim}}{2.5}}}
\end{split}
.\end{equation}
Finally, the noise level can be incorporated into each pixel of the galaxy image by sampling a Gaussian distribution with a mean of 0 and a standard deviation of $\sigma_e$ or $\sigma$, depending on the unit. This approach does not necessitate instrumental parameters such as dark current, $B_{\rm dark}$, and readout noise, $\sigma_{\rm RN}$, thereby offering a straightforward method to estimate noise levels when certain instrumental parameters are unavailable, particularly for planning instruments. In summary, this approach (or the one mentioned in Section~\ref{sec: postprocess}) can both be employed when adding noise to galaxy images, considering the availability of certain parameters.}

\end{appendix}

\end{document}